\documentstyle[prl,aps,floats,epsf,epsfig]{revtex}

\begin{document}

\draft

\title{$O(n)$ Spin Systems- Some General Properties: A Generalized 
Mermin-Wagner-Coleman Theorem, 
Ground States, Peierls Bounds,
and Dynamics}

\author{Zohar Nussinov}
\address{Institute Lorentz for Theoretical Physics, Leiden University\\
P.O.B. 9506, 2300 RA Leiden, The Netherlands}
\date{\today ; E-mail:zohar@lorentz.leidenuniv.nl}

\twocolumn[

\widetext
\begin{@twocolumnfalse}

\maketitle

\begin{abstract}

Here we examine classical $O(n)$ spin
systems with arbitrary two 
spin interactions (of unspecified
range) within a 
general framework. We 
shall focus on  
translationally invariant
interactions. In the this
case, we determine the 
ground states of the 
$O(n \ge 2)$ systems.
We further illustrate how one may 
establish Peierls bounds
for many Ising systems
with long range interactions.
We study the effect of thermal 
fluctuations on the ground states
and derive the corresponding fluctuation
integrals. The study of the thermal 
fluctuation spectra will lead us to
discover a very interesting
odd-even $n$ (coupling-decoupling)
effect. We will prove a generalized 
Mermin-Wagner-Coleman theorem
for all two dimensional systems 
(of arbitrary range) 
with analytic kernels
in $k$ space.
We will show that many three dimensional
systems have smectic like thermodynamics.
We will examine the topology of 
the ground state manifolds
for both translationally
invariant and spin glass
systems. We conclude with a 
discussion of $O(n)$ 
spin dynamics in the 
general case.

\end{abstract}

\vspace{0.5cm}

\narrowtext

\end{@twocolumnfalse}
]

\section{Introduction}

In this article we aim to unveil some of general
properties of $O(n)$ spins systems having 
two-spin interactions. We shall mostly
concern ourselves with
translationally invariant
systems.

The outline is as follows:
In section (\ref{toy})
we introduce frustrated toy models 
rich enough to 
illustrate certain of the 
general features that 
we aim to highlight.
These models will be employed
for illustrative purposes only.
The bored reader is encouraged
to skip the somewhat verbose
exposition and merely
study the two Hamiltonians.

In section(\ref{Ising-gs}) we 
discuss the ground states of
Ising spin models and show
what patterns one should expect
in general. Once the ground states
will be touched on, we will head
on to show how Peierls bounds
may be established for many
systems having infinite range 
interactions if the 
ground states are simple. 
In section(\ref{ISING-MFT})
we shortly review mean field
solutions of the general two 
spin Ising models.

In section(\ref{n-gs}) 
we prove that, sans 
special commensurability effects, the 
ground states of all
$O(n \ge 2)$ will typically have 
a spiral like 
structure.

In section(\ref{s-stiffness}) we will
make an exceedingly simple spin 
wave stiffness analysis
to gauge the effect of thermal
fluctuations on the various 
$O(n \ge 2)$ ground states.

In section(\ref{XY-fluct}) we will  discus
thermal fluctuations within the framework
of ``soft-spin'' XY model. We will
see that the normalization constraint
gives a Dirac like equation. In the  
aftermath, the fluctuation spectrum will
be seen to match with that derived 
in section(\ref{s-stiffness}). We will
show possible links to smectic like
behavior in three dimensions.

Next, we go one step further to study the 
fully constrained ``hard-spin'' $O(2)$ and $O(3)$
models and show (in section \ref{M-W-low})
that {\bf all} translationally
invariant systems in two dimensions
with an analytic interaction kernel
never develop spontaneous magnetization.
At the end of the section our analysis
will match that of sections(\ref{s-stiffness}) and (\ref{XY-fluct})

We extend the Mermin-Wagner-Coleman theorem 
to {\bf all} two dimensional interactions
with an analytic interaction
kernel in momentum space

In section(\ref{n=3}) 
we will examine the ``soft-spin''
version of Heisenberg spins.
We will see that it might be
naively expected that the 
spin fluctuations in 
odd $n$ spin systems are 
larger than in those 
with an even number of
spin components.
One of the fluctuating 
spin components
will remain unpaired.

Next, in section(\ref{n=4}), 
we carry out the spin fluctuation
analysis for four component
soft spins to see that their
spectra coincides with that
predicted in the earlier spin
stiffness analysis.

In section(\ref{high_mft}) 
we compute the critical
temperature of all 
translationally
invariant $O(n \ge 2)$ 
spin models with 
mean field theory.

In section(\ref{spherical})
we show that in the limit
of large $n$ both odd 
and even component spin
systems behave in the same
manner. Essentially, they all
tend towards an ``odd'' behavior.  

In section(\ref{brilliant}) 
we briefly remark that much of
analysis is not changed for 
arbitrary two spin interactions
(including spin glass models).

We conclude with a discussion
of $O(n)$ spin dynamics.

A central theme which will be repeatedly 
touched on throughout the paper
is the possibilities of non-trivial
ground state manifolds. If the 
system is degenerate the effective 
topology of the low temperature
phase of the system may be classified
in momentum (or other basis). In such
instances the low temperature behavior
of the systems will be exceedingly
rich.

\section{Definitions}
\label{definitions}

We will consider simple classical 
spin models of the type

\begin{eqnarray}
H = \frac{1}{2} \sum_{\vec{x},\vec{y}}
\hat{V}(\vec{x},\vec{y})[\vec{S}(\vec{x}) 
\cdot \vec{S}(\vec{y})]. 
\end{eqnarray} 

Here, the sites $\vec{x}$ and $\vec{y}$
lie on a (generally hypercubic)  
lattice of size $N$. The spins $\{ S(\vec{x}) \}$
are normalized and have $n$ components:

\begin{eqnarray}
\sum_{i=1}^{n} S_{i}^{2}(\vec{x}) = 1
\end{eqnarray}
at all lattice sites $\vec{x}$.

We shall, for the most part,
consider translationally 
invariant interactions $V(\vec{x},\vec{y}) 
= V(\vec{x}- \vec{y})$. 

We employ the non-symmetrical
Fourier basis convention

($f(\vec{k}) = \sum _{\vec{x}} 
F(\vec{x}) e^{-i \vec{k} \cdot \vec{x}};$
\bigskip
$ ~ F(\vec{x}) = \frac{1}{N} \sum_{\vec{k}} 
 f(\vec{k}) e^{i \vec{k} \cdot \vec{x}}$) 
wherein the Hamiltonian is diagonal and reads 

\begin{eqnarray}
H =  \frac{1}{2N} \sum_{\vec{k}} v(\vec{k}) |\vec{S}(\vec{k})|^{2}
\end{eqnarray}

where $v(\vec{k})$ and $\vec{S}(\vec{k})$ are the Fourier
transforms of $V(\vec{x})$ and $\vec{S}(\vec{x})$.

More generally, for some of the 
properties that we will illustrate,
one could consider any arbitrary real 
two spin interactions $\langle \vec{x} | V | \vec{y} \rangle$
which would be diagonalized in another basis 
$\{|\vec{u} \rangle\}$ instead of the Fourier 
basis. 

For simplicity, we will set the lattice constant to unity-
i.e.  on a hypercubic lattice (of side $L$) with
periodic boundary conditions the wave-vector
components $k_{l} = \frac{2 \pi r_{l}}{L}$
where $r_{l}$ is an integer (and the real space 
coordinates  $x_{l}$ are integers).

Throughout this work we will use 
$\vec{K}$ to denote reciprocal 
lattice vectors and
$\Delta (\vec{k})$  as a shorthand 
for the lattice lattice Laplacian:

\begin{eqnarray} 
\Delta(\vec{k}) = \sum_{l=1}^{d} (1-\cos k_{l}). 
\end{eqnarray}

In some of the frustrated systems that we will soon consider,  
$v(\vec{k})$ may be written explicitly as 
the sum of several terms: those favoring 
homogeneous states $\vec{k} \rightarrow 0$, 
and those favoring zero wavelength 
$\vec{k} \rightarrow \infty$ 
(or $\vec{k} \rightarrow (\pi,\pi,...,\pi)$
on a lattice.) As a result of this competition,
modulated structures arise on an intermediate scale.

\section{Toy Models- For illustrative purposes only}
\label{toy}

Although we will keep the discussion 
very general, it might be useful to have 
a few explicit applications in mind.
There is a lot of physical intuition 
which underlies the upcoming models. Unfortunately,
insofar as we are concerned, they will
merely serve as nontrivial toy models 
on which we will able to exercise 
our newly gained intuition.

The systems to be presented
are frustrated: not all two
spin interactions can be 
simultaneously satisfied. 

We choose these rather 
nontrivial toy examples
as they highlight some 
possible richness 
which is typically 
absent in the more
standard spin models.
As such, they will 
point to typically 
ignored subtleties.

\bigskip

\bigskip

{\bf{The Coulomb Frustrated Ferromagnet}}

Let us introduce our first toy
model ``The Coulomb Frustrated Ferromagnet''.

This is a toy model of a doped Mott insulator, 
where the tendency, of holes, to phase
separate at low doping is  
frustrated, in part, by electrostatic repulsion \cite{steve}.
In three dimensions, a simple spin Hamiltonian \cite{us}
which represents these 
competing interactions is 
\begin{eqnarray}
H_{Mott} = - \sum_{\langle \vec{x},\vec{y} \rangle} S(\vec{x}) S(\vec{y}) +  
\frac{Q}{8 \pi} \sum_{\vec{x} \neq \vec{y}} \frac{S(\vec{x}) 
S(\vec{y})}{|\vec{x}-\vec{y}|} \nonumber
\\ = \frac{1}{2N} 
\sum_{\vec{k}} [\Delta(\vec{k})+ \sum_{K} |\vec{k}-\vec{K}|^{-2}]
 |S(\vec{k})|^{2}. 
\end{eqnarray}
Here, $S(\vec{x})$ is  a coarse grained scalar variable which represents the 
local density of mobile holes. Each site $\vec{x}$ 
represents a small region of space in which
$S(\vec{x})>0$, and $S(\vec{x})<0$ correspond to hole-rich and 
hole-poor
phases respectively.  
In this Hamiltonian, the first ``ferromagnetic''
term represents the short-range
(nearest-neighbor) tendency of the holes to phase-separate and form 
a hole-rich ``metallic'' phase, whereas the frustrating effect of the 
electrostatic repulsion between holes is present in the second term.
Non-linear terms in the full Hamiltonian typically fix the locally
preferred values of $S(\vec{x})$. One may consider
$d \neq 3 $ dimensional variants wherein the spins 
lie on a hypercubic lattice, and the Coulomb kernel
in $H_{0}$ is replaced by $\frac{Q}{2 \Omega_{d}}|\vec{x}-\vec{y}|^{2-d}$ 
(or by $[\frac{Q}{4 \pi}~ \ln |\vec{x}- \vec{y}|~]$ in two dimensions) where
$\Omega_{d} = 2 \pi^{d/2}/\Gamma (d/2)$.
Here the competition between both terms, when $Q \ll 1$ 
favors states with wave-numbers $\simeq  Q^{1/4}$.
The introduction of the Coulomb interaction is brute force
non-perturbative: it is long range. Moreover,
the previous ferromagnetic ground state becomes, 
tout a' coup, infinite in energy. 
We will, for the large part, focus on the 
continuum limit of this Hamiltonian 
where the kernel
becomes
\begin{eqnarray}
v_{Mott}^{cont}(\vec{k}) = Q k^{-2}+ k^{2}[1+\sum_{\vec{K} \neq 0} 
(\frac{4 K_{l}^{2}}{K^{6}}-\frac{1}{K^{4}})]
\end{eqnarray} 
After rescaling,
this may also be regarded as the small $\vec{k}$ 
limit of the more general 
\begin{eqnarray}
v_{Q}(\vec{k})= \Delta(\vec{k}) + Q [\Delta(\vec{k})]^{-1} +
A [\Delta(\vec{k})]^{2} \nonumber
\\ + \lambda \sum_{i \neq j} (1- \cos k_{i})(1-\cos k_{j}) + O(k^{6}).
\end{eqnarray}

The constants $A$ and $\lambda$ are pinned
down if we identify $v_{Q}(\vec{k})=v_{Mott}(\vec{k})$.
Here, we will modify them
in order to streamline the quintessential
physics of this system. First, we set $A=0$: In the continuum
limit this term is not large nor does it
lift the ``cubic rotational
symmetry'' of the lattice
(i.e. those transformations which 
leave $\Delta(\vec{k})$ invariant) present to lower 
order. Next, we allow $\lambda $ to vary in order to 
turn on and off ``cubic rotational symmetry''
breaking effects.

Note that $v_{Q}(\vec{k})$ with may be regarded as 
$v_{Mott}$ augmented by all possible next to nearest
neighbor interactions. As our Hamiltonian respects
the hypercubic $d=2$ point symmetry group,
by surveying all possible
values of $\lambda$ 
we should be able to make
general statements regarding the
possible phases
(within the planes) 
of real doped Mott insulators. When $\lambda >0$
the minimizing wave-vectors will lie along the 
cubic axis and ``horizontal''  order
will be expected. When $\lambda <0$ 
the minimizing wave-vectors lie along
the principal diagonals and diagonal order 
is expected. At large values of $Q$,
when the continuum limit no longer
applies, trivial extensions of
these minimizing modes are
encountered where one or more of
the wave vector components is
set to $\pi$.

Note that if the ferromagnetic system were 
frustrated by a general long range kernel
of the form  $V(|\vec{x}-\vec{y}|) \sim |\vec{x}-\vec{y}|^{-p}$
we could replace the $[\Delta(\vec{k})]^{-1}]$ in $v_{Q}(\vec{k})$
by the more general $[\Delta(\vec{k})]^{(p-d)/2}$.
Here,  in the continuum limit, the minimizing modes
 are $\sim Q^{1/(2+d-p)}$ and as 
the reader will later be able easily verify all 
our upcoming analysis can be reproduced
for any generic long range frustrating interactions 
with identical conclusions.

\begin{figure}
\centering
\hspace{0.0in}{\psfig{figure=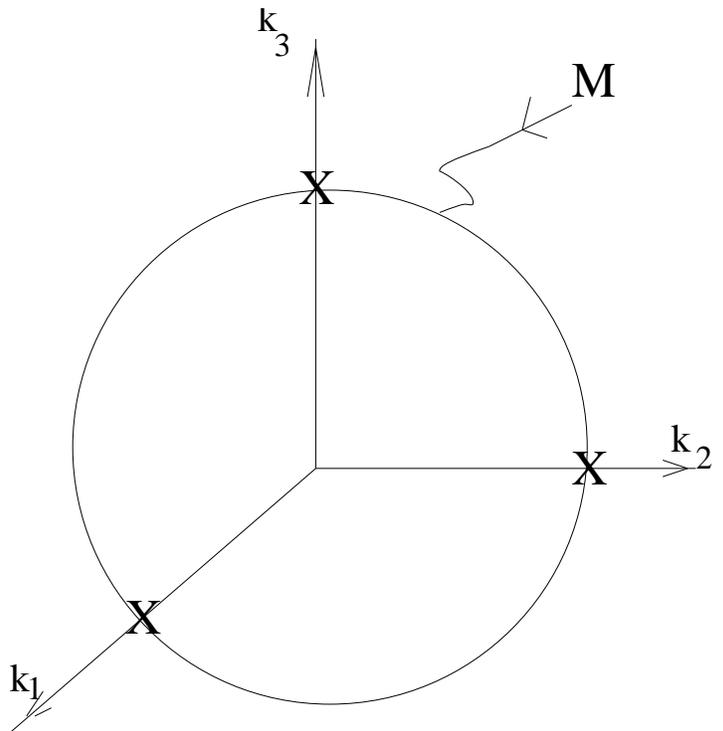,width=3.7in,clip=}}
\caption{The shell of modes which minimize the 
energy in the continuum limit. For the case just discussed
This sphere is of radius $|\vec{q}|=Q^{1/4}$.}
\label{fig:Minimizing_manifold_1}
\end{figure}

Shown above is the manifold ($M$) of the minimizing modes
in $\vec{k}$ space. When no symmetry
breaking terms ($ \lambda =0$) are present,
in the continuum
limit it is $M$ is the surface of sphere of radius $Q^{1/4}$.
If $\lambda \neq 0$ 
this degeneracy will be lifted:
only a finite number of modes will minimize the energy.
When $\lambda >0$ there will be $2d$ minimizing modes (denoted by the big X
in the figure) along the coordinate axes. In the 
up and coming we will focus mainly
on $ \lambda \ge 0$.
When $\lambda <0$, a moment's
reflection reveals that there will be $2^{d}$
minimizing modes along the diagonals,
i.e. parallel to $(\pm 1,\pm 1,\pm 1)$
(and in this case, they will have 
a modulus which differs 
from $Q^{1/4}$).

Unless explicitly stated otherwise, 
we will set $\lambda=0$
for calculational convenience
and when a finite $\lambda$ is
invoked it will be made 
positive (to avoid the $\lambda$ 
dependence of $|\vec{q}|$ incurred
when the former is negative).
At times, we will present results
for $v_{Q}(\vec{k})$ at sizable $Q$,
even though the model was motivated as
a good caricature of $v_{Mott}$
only in the continuum limit (at 
small wave-vectors  $ \sim Q^{1/4}$).

{\bf{Membranes}}

In several
fluctuating membrane systems, 
the affinity of the molecular 
constituents (say A and B) 
for regions of different
local curvature frustrates
phase separation \cite{Seul}.

Let us define $S(\vec{x})$ to be the difference between 
the A and B densities at $\vec{x}$. 

In the continuum, the energy of the system contains a contribution,
\begin{eqnarray}
H_{mix} = \frac{b}{2} \int d^{2}x~|\nabla S|^{2} 
\end{eqnarray}
reflecting the demixing of A and B species.
Instead of considering long-range interactions,
we now allow for out-of-plane (bending) distortions of the sheet. 
Specifically, we assume that the two molecular 
constituents display an affinity
for regions of different local curvature of the sheet. 
This tendency can be modeled by introducing a 
coupling term between the local composition $S(\vec{x})$
and the curvature of the sheet.

Provided that the distortions remain small,
we may write 
\begin{eqnarray}
H_{c} = \int d^{2}x~ [\frac{1}{2} \sigma
 |\nabla h(\vec{x})|^{2}+\frac{\kappa}{2}[\nabla^{2} h(\vec{x})]^{2}+
\Lambda S(
\vec{x}) \nabla^{2} h(\vec{x})]
\nonumber
\\
\equiv \int d^{2}x {\cal{H}}_{c},
\nonumber
\end{eqnarray}
where $h(\vec{x})$ represents the height profile of the sheet
(relative to a flat reference state),
$\sigma$ is its surface tension , and $\kappa$ 
is its bending modulus; $\Lambda$, the coefficient of the 
last term in the expression measures the strength of the coupling 
of the local curvature $\nabla^{2}h$ and the local composition 
$\phi$, which we have included here to lowest (bilinear) order.
This coupling term reflects the different affinities of the molecular 
constituents A $(S =1$ corresponds to pure A composition)
and B ($S =0$ corresponds to pure B composition) for, respectively,
 convex ($\nabla^{2}h>0$)
and concave ($\nabla^{2}h<0$) regions of the interface. 
We now minimize the total energy $H=H_{\phi}+H_{c}$,
w.r.t the membrane shape $\{h(\vec{x})\}$.

\begin{eqnarray}
0= \delta H_{c}= \int ~d^{2} x~ [\frac{\partial{\cal{H}}}{\partial h}
 \delta h+ \frac{\partial {\cal{H}}_{c}}{\partial (\partial_{i} h)}
 \delta( \partial_{i}h) +
\frac{\partial {\cal{H}}_{c}}{\partial(\partial_{i}^{2}h)}
 \delta(\partial_{i}^{2}h)] \nonumber
\\ = \int ~d^{2}x~[\frac{\partial {\cal{H}}_{c}}{\partial h}-
\partial_{i} \frac{\partial {\cal{H}}_{c}}{\partial (\partial_{i} h)} 
+\partial_{i}^{2} \frac{\partial {\cal{H}}_{c}}
{\partial (\partial_{i}^{2} h)}] \delta h(\vec{x}),
\nonumber
\end{eqnarray} 
the variational eqns,
where in obtaining to the last line we have employed
$\delta(\partial_{i}^{2} h(\vec{x})) = \partial_{i}^{2} \delta h(\vec{x})$,
$\delta(\partial_{i} h(\vec{x})) =\partial_{i} \delta h(\vec{x})$,
and integrated by parts twice. 

Thus
\begin{eqnarray}
-\sigma \nabla^{2} h + \kappa \nabla^{2} (\nabla^{2} h)+
 \Lambda \nabla^{2} S =0.
\end{eqnarray}
If $|\kappa \nabla^{2}(\nabla^{2} h)| \ll
 \min \{ |\sigma \nabla^{2} h|,|\Lambda \nabla^{2} S| \}$ 
then an approximate solution to the last eqn is
\begin{eqnarray}
\Lambda S \approx \sigma h,
\end{eqnarray}
and in ${\cal{H}}_{c}$, after an integration by parts,
\begin{eqnarray}
\Lambda S \nabla^{2} h \approx \Lambda S
 \frac{\Lambda}{\sigma} \nabla^{2} S
\rightarrow -\frac{\Lambda^{2}}{\sigma}(\nabla S)^{2}.
\end{eqnarray}
\begin{eqnarray}
H_{mix}+H_{c} \approx \int d^{2}x~[\frac{1}{2} b^{\prime} |\nabla S|^{2} +
 \frac{\Lambda^{2} \kappa}{2 \sigma^{2}} (\nabla^{2}S)^{2}] \nonumber
\\ \mbox{where}~~b^{\prime} \equiv b-\frac{\Lambda^{2}}{\sigma}.
\end{eqnarray}

This effective energy reads
\begin{eqnarray}
H = \int d^{2}k~ v_{membrane}(\vec{k}) |\phi(\vec{k})|^{2}, 
\end{eqnarray}
where $v_{membrane}(\vec{k}) = 
\frac{b^{\prime}}{2}k^{2}+\frac{\Lambda^{2} \kappa}
{2 \sigma^{2}}k^{4}$ is the 2D Fourier transform. 
A negative $b^{\prime}$ obtained when $b<\Lambda^{2}/\sigma$,
signals the onset of a curvature instability of the sheet.
This instability generates a pattern of domains that differ in
composition as well as in local curvature and thus assume 
convex or concave shapes.
The characteristic domain size corresponds to the existence of 
the minimum of the free energy at a non-zero wave number.
The modulation length $d \simeq
\sqrt{(\Lambda^{2} \kappa/\sigma^{2})/|b^{\prime}|}$.

After scaling, this model may be regarded as the continuum version of the 
frustrated short
range kernel
\begin{equation}
  v_{z}(\vec{k}) = z \Delta^{2}(\vec{k}) - \Delta (\vec{k})
\end{equation}
(where $z= -\Lambda^{2}\kappa /(\sigma^{2} b^{\prime})$)
on the lattice.

The real lattice Laplacian
\begin{equation}
  \langle \vec{x}| \Delta |\vec{y} \rangle  = \left\{ \begin{array}{ll}
      2d & \mbox{ for $\vec{x}=\vec{y}$} \\
      -1 & \mbox{ for $||\vec{x}-\vec{y}||_{\infty} = 1$}
\end{array}
\right.
\end{equation}

Notice that  $\langle \vec{x}| \Delta^{R} |\vec{y} \rangle = 0 
\mbox{ for $ ||\vec{x}-\vec{y} ||_{\infty} > R$ (= Range) }$.
Our system is of $Range=2$.

Explicitly
\begin{eqnarray}
  \langle \vec{x}| \Delta^{2}| \vec{y} \rangle \mbox{ } = \mbox{ }&& 2d(2d+2) \mbox{
    for } \vec{x} = \vec{y} \nonumber
  \\
  \ \ && -4d \mbox{ for  } |\vec{x}-\vec{y}| = 1 \nonumber \\
  \ \ && 2 \mbox{ for } (\vec{x} -\vec{y}) = 
(\pm \hat{e}_{\ell} \pm \hat{e}_{\ell^{\prime}}) \mbox{ where  }
 \ell \neq \ell^{\prime}      \nonumber \\
  \ \ && 1 \mbox{ for a $ \pm 2 \hat{e}_{\ell} $ separation}.
\end{eqnarray}

We shall extend the investigation of
this model over a broader range
of parameters than suggested 
by its initial physical motivation.

Note that, in the continuum limit, theories 
with high order derivative terms
will generally give rise to 
\begin{eqnarray}
v(\vec{k})  = P(k^{2})
\end{eqnarray} 
where $P$ is some polynomial.
Although $v_{z}(\vec{k})$ and
its likes are  artificial
on the lattice, their continuum 
limit is quite generic.
Later on  we will show that if 
$P(k^{2})$ attains its global
minima at finite $|\vec{k}|$,
then thermal instabilities
can incur an extremely 
low value of $T_{c}$.

\section{Ising Ground States}
\label{Ising-gs}

In an ``Ising''  system $S(\vec{x})= \pm 1$
everywhere on the lattice. Stated alternatively, 
the scalar ($n=1$) spins satisfy a normalization
constraints 
\begin{equation}
\{ S^{2}(\vec{x}) =1 \}
\end{equation}
at all $N$ lattice
sites $\vec{x}$. Henceforth, we will adopt
the latter point of view. 

Let us define the manifold $M$ spanned by the set of minimizing wave-vectors
 $\vec{q}$ 
\begin{eqnarray}
v(\vec{q} \in M) \equiv \min_{\vec{k}} \{ v(\vec{k}) \}.
\end{eqnarray}

If the local normalization constraints are swept
aside then it is clear that the ground states
are superpositions of sinusoidal waves with
wave-vectors $\vec{q} \in M$. One would 
expect this to be true, in spirit, also in
the highly constrained Ising case,
if $v(\vec{k})$ is sharply dipped at 
its global minima.  ``Digitizing'' 
a particular plane wave 
\begin{eqnarray}
S(\vec{x}) = sign(\cos(\vec{q}_{1} \cdot \vec{x}))
\end{eqnarray}
and comparing it with the exact (numerical) 
ground state, one finds encouraging  agreement
in certain cases. 

For instance, this gives reasonable
accord when $H= H^{Mott}$. 

This Hamiltonian (with some twists) was investigated
in \cite{steve}
on a square ($d=2$) lattice.

Note that in the continuum limit 
(i.e. if the lattice is thrown
away) we might naively anticipate a huge
ground state degeneracy- 
a ``digitized plane wave'' 
for each wave vector $\vec{q}$ lying on
the $(d-1)$ dimensional manifold
 $\{ M_{Q} : q^{4} = Q \}$
This large degeneracy 
might give rise to
a loss of stability 
against thermal fluctuations.

It is found that striped phases
(i.e. ``digitized plane waves'') were found
in virtually all
of the parameter range.
Only for a very small range of parameters
were more complicated periodic structures
found.

An intuitive feeling can be gained by considering 
a one dimensional pattern such as 
\begin{eqnarray}
++--++--++--...
\end{eqnarray}
This pattern is
a pure mode 
\begin{eqnarray} 
S_{period=4}(x) = \sqrt{2} \cos[\frac{\pi}{2} x - \frac{\pi}{4}]. 
\end{eqnarray}
A double checkerboard pattern such as 
\begin{eqnarray}
 ++--++-- \nonumber
\\ ++--++-- \nonumber
\\ --++--++ \nonumber
\\ --++--++ 
\end{eqnarray}
extending in all directions in the plane is thus trivially given by
\begin{eqnarray}
S(\vec{x}) = 2 \cos[\frac{\pi}{2} x_{1} - 
\frac{\pi}{4}] \cos[\frac{\pi}{2} x_{2} - \frac{\pi}{4}] = \nonumber
\\  \cos[\frac{\pi}{2}(x_{1}+x_{2})-\frac{\pi}{2}] +
 \cos[\frac{\pi}{2}(x_{1}-x_{2})].
 \end{eqnarray}
Such a $ 4 \times 4 \times 4$ periodic pattern in 
three dimensions would include the eight modes 
$\frac{1}{2} (\pm \pi,\pm \pi . \pm \pi)$. 
This trivial example serves to illustrate an simple point. 
If one has a periodic building block
of dimensions $p_{1} \times p_{2} \times p_{3}$, then 
\begin{eqnarray}
S(\vec{x}) = S_{p_{1}}(x_{1}) S_{p_{2}}(x_{2}) S_{p_{3}}(x_{3}).  
\end{eqnarray}
If a configuration $S_{p}(x)$ contains the modes $\{ k_{p}^{m} \}$
with amplitudes $\{ S_{p}(k_{m}) \}$, then Fourier transforming 
the periodic configuration $S(\vec{x})$ one will find  the modes
$(\pm k_{p_{1}}^{m_{1}}, \pm k_{p_{2}}^{m_{2}}, \pm k_{p_{3}}^{m_{3}})$
appearing with a weight $ \sim |S_{p_{1}}(k_{1}) \times S_{p_{2}}(k_{2})
\times S_{p_{3}}(k_{3})|^{2}$. For high values of the periods $p$, 
the weight gets scattered over a large set of wave-vectors. 
If $v(\vec{k})$ has sharp minima, such states will not be favored.
The system will prefer to generate patterns s.t. in all 
directions $i$ albeit one $p_{i}=1 ~~(\mbox{or perhaps~ }2)$. For a 
$p_{1} \times p_{2} \times p_{3}$ repetitive pattern, the discrete
Fourier Transform will be nonzero for only $\Pi_{i=1}^{3} ~p_{i}$
values of $\vec{k}$. This trivial observation  suggests
the phase diagram obtained by U. Low et al.\cite{Low} in the two 
dimensional case. The intuition is obvious.
We have derived \cite{us},  rigorously, the ground states
in only several regions of its parameter space
(those corresponding to ordering with half a reciprocal
lattice vector),  and on  a few 
special surfaces (corresponding to ordering 
with a quarter of a reciprocal lattice vector).
In all of these cases the Ising states may
be expressed as superpositions of the
lowest energy modes
$\exp[i \vec{q}_{m} \cdot \vec{x}]$.
Lately, a beautiful extension was carried out
by \cite{Gilles}.

We now ask whether commensurate lock-in is to be expected.
The energy of the Ising ``digitized plane wave''
on an $L \times L \times L$ lattice where
$\vec{q} = (q_{1},0,0)$ with $q_{1} = 2 \pi/m$, with even $m$, reads
\begin{eqnarray}
E = \frac{1}{2N} \sum_{\vec{k}} v(\vec{k}) |S(\vec{k})|^{2} =
\nonumber
\\ \frac{8}{m} \sum_{j=1,3,...,m-1}
 \frac{v(\vec{k}=(\frac{2 \pi j}{m},0,0))}{|\exp[2 \pi i j/m]-1|^{2}} =
\nonumber
\\  \frac{2}{m} 
\sum_{j=1,3,...,m-1}
\frac{v(\vec{k}=( \frac{2 \pi  j}{m},0,0))}{\sin^{2}(\pi j/m)}. 
\end{eqnarray}
The lowest energy state amongst all
states of the form considered 
is a possible candidate
for the ground state. 

For the particular model long-range introduced above, 
it seems that for small values of $Q$, it might be 
worthwhile to have an incommensurate phase. This is,
in a sense, obvious- all low energy modes are of very small
wave-number and hence not of low commensurability. 
The energy
\begin{eqnarray}
E = \frac{1}{N} \sum_{n=0}^{\infty}
 \frac{16v(\vec{k}=(2n+1)\vec{q})}
{(2n+1)^{2}
\pi^{2}}.
\end{eqnarray} 
For $q \sim Q^{1/(d+1)} \ll 1$, the higher harmonics
$\vec{k} = (2n+1) \vec{q}$, do not entail high energies. 
For large values of $Q$, $q  \simeq O(1)$,
and $v(\vec{k}=(2n+1)\vec{q})$ can be very large if $[(2n+1)\vec{q}]$
approaches a reciprocal lattice vector $\vec{K}$.
Under these circumstances
it will pay off to have a commensurate structure; for a 
$u_{1} \times u_{2} \times ... \times u_{d}$ repetitive block only the modes 
$\vec{k} = 2 \pi(\frac{n_{1}}{u_{1}},\frac{n_{2}}{u_{2}},...,
\frac{n_{d}}{u_{d}})$
will be populated (i.e. have a non-vanishing
$|\vec{S}(\vec{k})|^{2}$)- the ferromagnetic point [a reciprocal lattice point]
will not be approached arbitrarily close- if that is not true
weight will be smeared over energetic modes. Generically, we will
not be expect commensurate lock-in in $\lim \vec{q} \rightarrow 0$
for {\it{any}} theory with a  frustrating long range
interaction. Although we have considered only 
striped phases (which have previously argued are the only
ones generically expected), it is clear that this 
argument may be reproduced for more exotic 
configurations.

For finite range interactions it is easy to
prove, by covering the system with large maximally
overlapping blocks, that there will be a 
sliver about $\vec{q} = 0$, for which we
will find the ferromagnetic 
ground state.

A polynomial in $\Delta(\vec{k})$ 
will have its minima at $\Delta(\vec{k}) = const$, i.e.
on a (d-1) dimensional hypersurface(s) in $\vec{k}$-space or
at the (anti)ferromagnetic point.

The kernel $v_{z}(\vec{k})= z \Delta^{2} - \Delta$
has its minima ($z > 0$) at
\begin{equation}
  \vec{q} \in M_{z} : \Delta (\vec{q}) = \min \{ \frac{1}{2z} \mbox{,
    } 4d \}
\end{equation}  
For $ z > \frac{1}{8d}$: $M_{z}$ is $(d-1)$ dimensional.

We may divide the lattice into all maximally over-lapping 
$5 \times 5 \times ...\times 5$
hyper-cubes centered about each site of the lattice.
\begin{equation}
  Energy = \frac{1}{5 \times 6^{d-1}} \sum_{hypercubes} \epsilon(hypercube)
\end{equation}
and evaluate the energies $\epsilon$ of all $5 \times 5 \times... \times 5$
Ising configurations. Of all $2^{5^{d}}$
configurations the Neel state will have the lowest energy for a
sliver about $z=\frac{1}{8d}$. Analogously for $z > z_{top}>>1$,
by explicit evaluation,  the ground state will
be ferromagnetic. Contour arguments can be employed
and a finite lower bound on $T_{c}$ generated.

In this system  on the square
lattice with periodic boundary 
conditions along all diagonals 
$ \hat{e}_{\pm} \mbox{ :}$ defined by $ x_{1} \mp x_{2} = const$ 
Ising ground states for
 $z = \frac{1}{8}$ can be synthesized.
All minimizing modes lie on
\begin{equation}
  \vec{q} \in M_{z=\frac{1}{8}} : |q_{1} \pm q_{2}| = \pi .
\end{equation} 
By prescribing an arbitrary spin configuration along $x_{-}$
\bigskip
 and fixing $S(x_{-},x_{+}) = S(x_{-},0) (-1)^{x_{+}}$:
\begin{equation}
  {S(\vec{k}) = \sum_{x_{-}} S(x_{-},0) \exp(ik_{-}x_{-}) \sum_{x_{+}}
    (-1)^{ x_{+}}\exp(ik_{+}x_{+})}
\end{equation}
vanishes for $|k_{+}| = |k_{1}+k_{2}| \neq \pi$.  Similarly by taking
the transpose of these configurations we can generate patterns having
$S(\vec{k}) = 0$ unless $|k_{-}| = |k_{1}-k_{2}| = \pi$.  The ground
state degeneracy is bounded from below by, the number of independent
spin configurations that can be fashioned along $x_{+} \mbox{ or }
x_{-} , (2^{L+1}-2)$ where $L$ is the length of the system along the
$x_{\pm}$ axis. The number of $\vec{q}$ values, commensurate with the
diagonal periodic boundary conditions, lying on $M_{z=\frac{1}{8}}$ is
$(4L-2)$.

\begin{figure}
\centering
\hspace{0.0in}{\psfig{figure=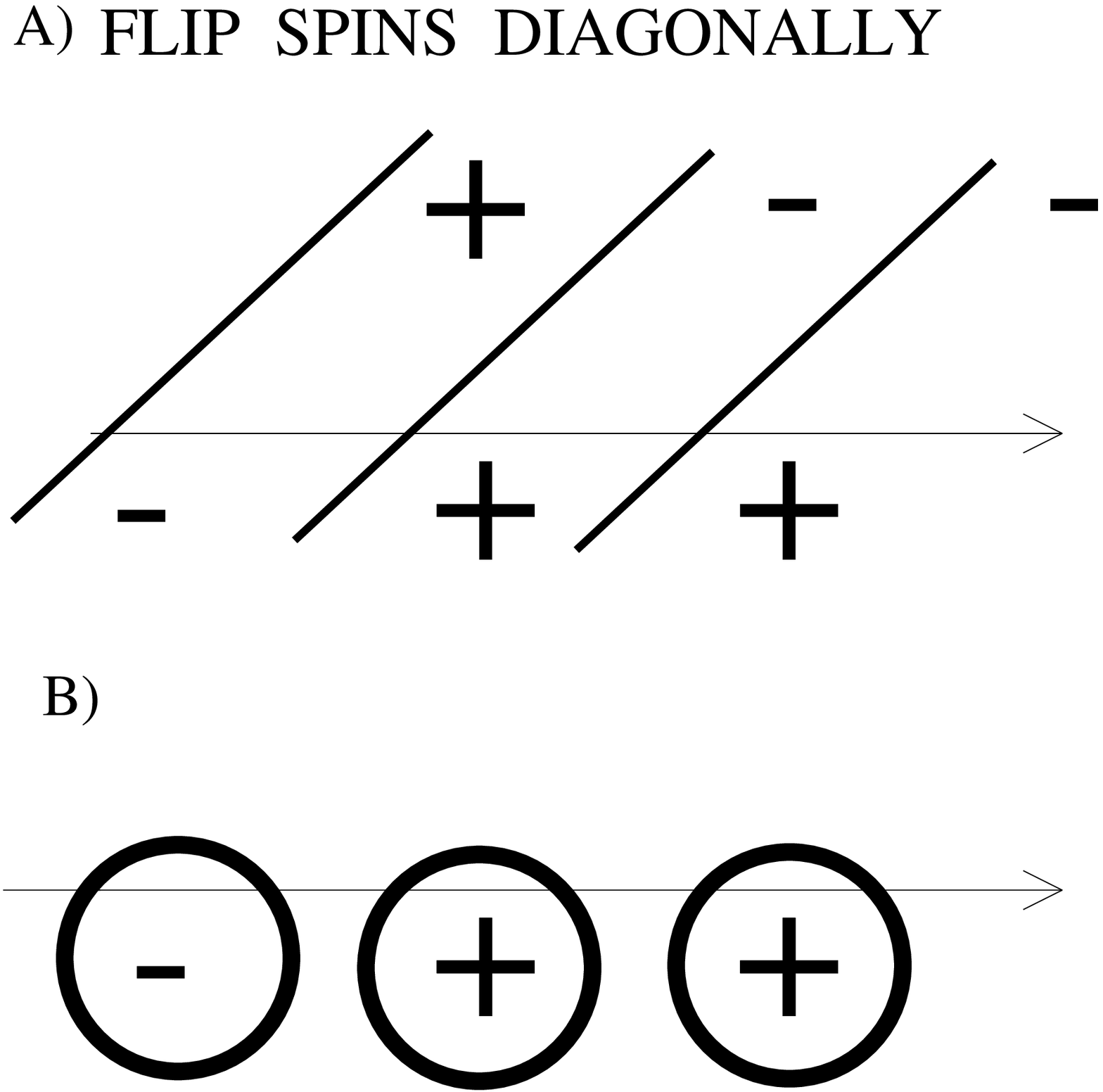,width=3.7in,clip=}}
\caption{Simple ground states for $v_{z}(\vec{k})$ with $z=1/8$
(and for $v_{Q}(\vec{k})$ when $Q=16$): A) Their construction- 
along the arrowed line we arbitrarily
prescribe spins. For each move along the dotted diagonal lines
we flip the spins. B) In this chain of
diagonal super-spins there
is  no stiffness against 
flipping.
}
\label{fig:diagonal_ground_states}
\end{figure}

[Similarly for $d>2$, one can set $(d-2)$ of the $\vec{q}$ components
to zero. There are $d(d-1)/2$ cross-sections of the d-dimensional
$M_{z=\frac{1}{8}}$, all looking like the the two-dimensional $M$ just
discussed (i.e.  $|q_{1} \pm q_{2}| = \pi$ ).  The real-space ground
state degeneracy is bounded from below by $d(d-1)[2^{L}-1]$ (along the
$(d-2)$ zero-mode directions the ground state spin configurations
display no flip).]

If we regard each diagonal row of
spins as a ``super-spin'' then  we will
see that flipping any ``super-spin'' entails
no energy cost. This is reminiscent to 
a nearest neighbor Ising chain where
the energy cost for flipping a spin
is dwarfed by comparison to the
(logarithmically) extensive 
entropy.
We might expect that here, too,
ordering might be somewhat inhibited.

In two dimensions 
\begin{equation}
  \lim _{z \rightarrow \infty} M_{z} : {\vec{q}}^{\mbox{ } 2} =
  \frac{1}{2z}
\end{equation}
the ``average'' number of allowed $\vec{q} \in M_{z} \mbox{ values }
\ll O(L)$ (and similarly for the onset $\lim_{z \rightarrow
  \frac{1}{8d}^{+}} M_{z} : (\vec{q}-(\pm \pi,\pm \pi))^{2} =
(\frac{1}{8d}-\frac{1}{2z})$).  For a hypercubic lattice of
size $L_{1} \times L_{2} \times ... \times L_{d}$
in  $d>2$  many discrete reciprocal points will give rise to the 
same value of $\Delta(\vec{k}) \sim \vec{k}^{2}$. The proof is trivial:
if all $L_{l} = L$, then the number of possible $\vec{k}^{2}$
values is bounded by $dL^{2}$, whereas there are $L^{d}$~ $\vec{k}-$values.

Therefore, on ``average'', the number of $\vec{k}$ points lying on $M_{z}$,
or more precisely lying  lying the closest to $M_{z}$, s.t. $|\Delta(\vec{k}) -
 \frac{1}{2z}|$
is min, is, at least, $O(L^{d-2})$. 

[Of these, $\frac{2^{d-z}~d!}{\Pi_{i}(n_{i}!)}$ wave-vectors,
 with $n_{i}$ (and $z$)
denoting the number of identical components (and the
number of zero components)
of a certain $\vec{k}_{1}$ nearest to $M_{z}$,
are related to $\vec{k}_{1}$ by symmetry.]

As we have stated previously, in the continuum limit ($q \rightarrow 0$)
any short range kernel (including this one) will have a uniform 
(ferromagnetic) ground state.
 However the impossibility
of constructing ground states that contain 
only ``good'' Fourier modes $S( \vec{k} \in M)$ when
$M$ shrinks to a curved surface enclosing the origin
is more general and will proved in the next section.

For this short range model, even
for $ z \neq \frac{1}{8}$
a  huge ground state degeneracy is expected. 
A ``plane wave''  might correspond to each
wave-vector $\vec{q}$ (or commensurate wave-vectors 
nearby) lying on the $(d-1)$ dimensional manifold $M$.

As we shall prove later on,  even in high dimensions, 
and even if the interactions are long ranged,
in the continuum limit it will not be possible 
to construct Ising states
in which $S (\vec{k} \not\in M) =0$ unless 
the minimizing manifold $M$ contains 
flat non-curved segments (or more 
generally intersects a plane
at many points).

\section{A universal Peierls bound}
\label{peierls}

If a real hermitian kernel $v(\vec{k})$ attains
its minima in only
a finite number of commensurate
reciprocal lattice points
$\{\vec{q}_{i}\}$,
then a Peierls bound can, in some instances,
be proven for an infinite range model:
When possible this is suggestive of a
finite $T_{c}$.  

For instance, the bound for a (lattice) Coulomb gas 
(with the kernel solving the
discrete Laplace equation on the lattice) is trivially
generated. 

\begin{eqnarray}
v_{e}(\vec{k}) - v_{e}(\vec{q}) = e/\Delta(\vec{k}) -
e/\Delta(\vec{q} = \pi,\pi,\pi) \nonumber
\\ \ge -A (\Delta(\vec{k}) -
\Delta(\vec{q} = \pi,\pi,\pi)) ),
\end{eqnarray}
with $A = 16 d^{2}$. The right hand side is the
kernel of an antiferromagnet. Both system share the same ground
states. For a given configuration the energy penalty for the Coulomb
gas 
\begin{eqnarray}
\Delta E_{e}= 1/(2N) \sum_{\vec{k}} [v_{e}(\vec{k})-v_{e}(\vec{q})]
|S(\vec{k})|^{2}
\end{eqnarray}
 is bounded from below by the corresponding penalty
in an antiferromagnet of strength $A$.  In $d=2$ the contour penalty
of the antiferromagnet is $ 2A|\Gamma|$, ($|\Gamma| \equiv$ length of
the contour $\Gamma$). A similar trick may frequently 
be employed when the minimizing wave-vectors attain other 
commensurate values. It relies on comparison to a short
range kernel for which a Peierls bound is trivial.

This can be extended quite generally.
All translationally invariant system
with commensurate minimizing 
wave-vectors 
($\vec{q} = (0,0, \ldots 0), (\pi, 0, \ldots 0), \ldots$)
at which the minimum of $v(\vec{k})$ at $\vec{k}=\vec{q}$
is quadratic may be bounded by a kernel 
of a system having nearest neighbor ferromagnetic 
and antiferromagnetic bonds for which the Peierls bound is trivial
(linear in the perimeter of the domain
wall $\Gamma$).

\section{Ising Weiss Mean-Field Theory}
\label{ISING-MFT}

Assuming that for $T<T_{c}$: $ \langle S(\vec{x}) \rangle = s~ sign[\cos(\vec{q} \cdot \vec{x})]$, 
then 
\begin{eqnarray}
\sum_{\vec{y}} \langle S(\vec{y}) \rangle V(\vec{x}=0,\vec{y}) =
\frac{1}{N} \sum_{\vec{k}} \langle S \rangle (\vec{k}) v(-\vec{k}) \nonumber \\
= s \sum_{n=0}^{\infty} \frac{4(-1)^{n}}{(2n+1)\pi}[v(\vec{k} = (2n+1) \vec{q})) + v(\vec{k} = -(2n+1) \vec{q})]
\end{eqnarray}
where we have assumed $q_{i} = \frac{t_{i}}{u_{i}}$ with $u_{i} \gg 1$ for all $d$ components in replacing a discrete Fourier transform sum
by an integral.
\begin{eqnarray}
s= \langle S(\vec{x}=0) \rangle = - \tanh[\beta \sum_{\vec{y}}
V(\vec{x}=0,\vec{y}) \langle S(\vec{y}) \rangle]
\end{eqnarray}
yields 
\begin{eqnarray}
\beta_{c}^{-1} = |\sum_{n=0}^{\infty} \frac{8(-1)^{n}}{(2n+1)\pi} v(\vec{k} = (2n+1) \vec{q})|.
\end{eqnarray}

The latter is an upper bound on $T_{c}$ as the non-trivial solution
[with a non-zero magnetization $s$] should be self-consistent at all
sites (not only at $\vec{x} =0$)

{\bf{Summary and Outlook}}

We have argued that for 
translation 
invariant two-spin 
kernels, the ground
states are generically 
uniform, Neel ordered, or 
thin stripes (all periodic
configurations in which
the net weight $\sum_{\vec{k}} |S(\vec{k})|^{2}$
is not smeared over many non-identical modes.)
If the Fourier transformed kernel
$v(\vec{k})$ is sharply peaked about
its global minima (at $\{\vec{q}\}$)
and if all other local minima
have a much higher ``energy''
$v(\vec{k})$  then the 
modulation length
is $O(|\vec{q}|^{-1})$.
All this might seem trivial/naive and 
incorrect. 
(It is also possible 
to get a feel for this 
via the examination 
of the local  magnetization
$ \langle S(\vec{x})) \rangle$ at low temperatures
as computed within the
dual unconstrained 
Hubbard Stratonovich
Hamiltonian.)

To emphasize that this is only generic
but non-universal, we have constructed
non-periodic ground states of a certain
finite ranged Hamiltonian.

As we have seen, frustrating
interactions can give rise to 
massive degeneracy. These could,
potentially, give rise to a
small value of $T_{c}$.

In the continuum limit
($|\vec{q}| \rightarrow 0$),
the ground states in
the presence of a 
frustrating long range
interaction are expected to
be non-commensurate;
if short range frustration
is present then the continuum limit
ground state will be homogeneous 
(ferromagnetic).

In this context, it might
be worthwhile to point out 
that the triangular antiferromagnet
was solved exactly by Wannier \cite{Wannier}. 
This system has finite entropy at $T=0$. There is no transition
at finite temperature- this, again, is presumably
related to this high ground state degeneracy.

\section{$O(n \ge 2)$ ground states}
\label{n-gs}

By an $O(n)$ system we mean that 
the spins $\{ \vec{S}(\vec{x}) \}$
have $n$ components and are all normalized to unity-
$\vec{S}^{2}(\vec{x})=1$.   

Our previous ansatz $S(\vec{x}) = sign(\cos(\vec{q}_{1} \cdot \vec{x}))$
is readily fortified in the
O($n \ge 2$) scenario: here there is no need to
``digitize''- in  the spiral

\begin{eqnarray}
S_{1}(\vec{x}) = \cos (\vec{q} \cdot \vec{x}) \nonumber
\\ S_{2}(\vec{x}) = \sin (\vec{q} \cdot \vec{x}) \nonumber
\\ S_{i>2}(\vec{x}) = 0
\end{eqnarray}

the only non-zero Fourier components are
$\vec{S}(\vec{q}) , \vec{S}(-\vec{q})$.
In plain terms, this state can be constructed
with the minimizing wave-vectors only.

It follows that any ground state $g$ must be of
the form
\begin{eqnarray}
S_{i}(\vec{x})= \sum_{m} a_{i}^{m} \cos(\vec{q}_{m} \cdot \vec{x} +
 \phi_{i}^{m}). 
\end{eqnarray}

Let us turn to the normalization of the spins at all sites. 

\begin{eqnarray}
1(\vec{x}) \equiv \sum_{i=1}^{n} S_{i}^{2}(\vec{x}) = \frac{1}{2} \sum_{m,m^{\prime}}
\sum_{i=1}^{n} a_{i}^{m} a_{i}^{m^{\prime}} 
\nonumber \\ (\cos(\phi_{i}^{m} + \phi_{i}^{m^{\prime}}) \cos[(\vec{q}_{m}+
\vec{q}_{m^{\prime}}) \cdot \vec{x}] \nonumber
\\ -\sin(\phi_{i}^{m}+\phi_{m^{\prime}})
\sin[(\vec{q}_{m}+\vec{q}_{m^{\prime}}) \cdot \vec{x}]  
\nonumber
\\ + \cos (\phi_{i}^{m} -\phi_{i}^{m^{\prime}}) \cos[(\vec{q}_{m} -
\vec{q}_{m^{\prime}}) \cdot \vec{x}] \nonumber 
\\ -\sin(\phi_{i}^{m}-\phi_{i}^{m^{\prime}})
\sin[(\vec{q}_{m} -\vec{q}_{m^{\prime}}) \cdot \vec{x}]) 
\end{eqnarray}

If $1(\vec{x})= 1$ is to hold identically for all sites $\vec{x}$,
 then all non-zero Fourier 
components must vanish. 

For the $\{ \cos(\vec{A} \cdot \vec{x})  \}$  Fourier components:

\begin{eqnarray}
0= [\sum_{\vec{q}_{m}+\vec{q}_{m^{\prime}}= \vec{A}}~~
 \sum_{i=1}^{n} a_{i}^{m} a_{i}^{m^{\prime}}
\cos (\phi_{i}^{m} + \phi_{i}^{m^{\prime}}) \nonumber \\
+ \sum_{\vec{q}_{m}-\vec{q}_{m^{\prime}} = \vec{A}}~~ \sum_{i=1}^{n} a_{i}^{m}
a_{i}^{m^{\prime}} \cos(\phi_{i}^{m}-\phi_{i}^{m^{\prime}})]
\end{eqnarray}

and a similar relation is to be satisfied by the $\{ \sin(\vec{A} \cdot
 \vec{x}) \}$ components:

\begin{eqnarray}
0= [\sum_{\vec{q}_{m}+\vec{q}_{m^{\prime}}= \vec{A}}~~ \sum_{i=1}^{n}
 a_{i}^{m} a_{i}^{m^{\prime}}
\sin (\phi_{i}^{m} + \phi_{i}^{m^{\prime}}) \nonumber \\
+ \sum_{\vec{q}_{m}-\vec{q}_{m^{\prime}} = \vec{A}}~~ \sum_{i=1}^{n}
a_{i}^{m} a_{i}^{m^{\prime}} \sin (\phi_{i}^{m}-\phi_{i}^{m^{\prime}})]
\end{eqnarray}

\begin{eqnarray}
a_{i}^{m} \cos \phi_{i}^{m} \equiv v_{i}^{m},~~ a_{i}^{m} \sin 
\phi_{i}^{m} \equiv u_{i}^{m}.
\end{eqnarray} 
The $\{ \cos(\vec{A} \cdot \vec{x}) \}$ and $\{ \sin(\vec{A} \cdot
 \vec{x}) \}$ conditions read 
\begin{eqnarray}
0= \sum_{\vec{q}_{m}+\vec{q}_{m^{\prime}}= \vec{A}}~~ \sum_{i=1}^{n}
[v_{i}^{m} v_{i}^{m^{\prime}} - u_{i}^{m} u_{i}^{m^{\prime}}]
\nonumber
\\ +\sum_{\vec{q}_{m}-\vec{q}_{m^{\prime}} = \vec{A}}  \sum_{i=1}^{n} 
[v_{i}^{m} v_{i}^{m^{\prime}} + u_{i}^{m} u_{i}^{m^{\prime}}] \nonumber
\\ 0= \sum_{\vec{q}_{m}+\vec{q}_{m^{\prime}}= \vec{A}}~~ \sum_{i=1}^{n} 
[v_{i}^{m} u_{i}^{m^{\prime}}+u_{i}^{m} v_{i}^{m^{\prime}}]\nonumber
\cr + \sum_{\vec{q}_{m}-\vec{q}_{m^{\prime} = \vec{A}}} ~~ \sum_{i=1}^{n}
[v_{i}^{m} u_{i}^{m^{\prime}} - u_{i}^{m} v_{i}^{m^{\prime}}] 
\end{eqnarray} 
In  the case of two pairs of wave-vectors $\pm \vec{q}_{1}$ and
$\pm \vec{q}_{2}$, both not equal half a reciprocal lattice vector:
$(0,0,..,), (\pi,0,...,0), ~(0,\pi,0,...,0)$,...,
$(\pi,\pi,0,...,0),...,(\pi, \pi,...,\pi)$,
the vector $\vec{A}$ (up to an irrelevant sign) may attain four non-zero values:
$\vec{A} = 2\vec{q}_{1},2\vec{q}_{2},\vec{q}_{1} \pm \vec{q}_{2}$.

When $\vec{A} = \vec{q}_{1} + \vec{q}_{2}$,~~ the conditions are
\begin{eqnarray}
0 = \sum_{i=1}^{n} [v_{i}^{1} v_{i}^{2} - u_{i}^{1} u_{i}^{2}], \nonumber
\\ 0= \sum_{i=1}^{n} [v_{i}^{1} u_{i}^{2} + u_{i}^{1} v_{i}^{2}].
\end{eqnarray} 
When $\vec{A} = \vec{q}_{1} - \vec{q}_{2}$, these conditions read
\begin{eqnarray}
0= \sum_{i=1}^{n} [v_{i}^{1} v_{i}^{2} + u_{i}^{1} u_{i}^{2}] \nonumber
\\ 0 = \sum_{i=1}^{n} [v_{i}^{1} u_{i}^{2} - u_{i}^{1} v_{i}^{2}]
\end{eqnarray}
For $\vec{A} = 2\vec{q}_{\alpha}$ ~($\alpha =1,2$):
\begin{eqnarray}
0= \sum_{i=1}^{n} [v_{i}^{\alpha} v_{i}^{\alpha} - u_{i}^{\alpha}
 u_{i}^{\alpha}]  \nonumber
\\ 0= 2 \sum_{i=1}^{n} u_{i}^{\alpha} v_{i}^{\alpha} 
\end{eqnarray}
Define 
\begin{eqnarray}
\vec{U}^{\alpha} \equiv (u_{i=1}^{\alpha},u_{2}^{\alpha},...,u_{n}^{\alpha})
 \nonumber
\\ \vec{V}^{\alpha} \equiv (v_{1}^{\alpha},...,v_{n}^{\alpha})
\end{eqnarray}
The previous conditions imply that 
\begin{eqnarray}
\vec{V}^{1} \cdot \vec{U}^{2} = \vec{U}^{1} \cdot \vec{V}^{2} =0 \nonumber
\\ \vec{V}^{1} \cdot \vec{V}^{2} = \vec{U}^{1} \cdot \vec{U}^{2} =0 \nonumber
\\ \vec{U}^{1} \cdot \vec{V}^{1} = \vec{U}^{2} \cdot \vec{V}^{2} =0. 
\label{ortho}
\end{eqnarray}

The four vectors $\{ \vec{U}^{1},\vec{U}^{2},\vec{V}^{1},\vec{V}^{2} \}$
are all mutually orthogonal.  The number of spin components $n \ge 4$.

Two additional demands that follow are
\begin{eqnarray}
\vec{V}^{\alpha} \cdot \vec{V}^{\alpha} = \vec{U}^{\alpha} \cdot
 \vec{U}^{\alpha} \nonumber
\\ \sum_{\alpha=1}^{2} [\vec{V}^{\alpha} \cdot \vec{V}^{\alpha} +
 \vec{U}^{\alpha} \cdot \vec{U}^{\alpha}] = 2 \sum_{\alpha} \vec{V}^{\alpha}
 \cdot \vec{V}^{\alpha} = 2.
\label{norm}
\end{eqnarray}
The last equation is the normalization condition- the statement that the
 coefficient
of $\cos (\vec{A} \cdot \vec{x})$, when $\vec{A} =0$, is equal to 1. 

For the case of a single pair of wave-vectors, $\pm \vec{q}_{1}$, 
~~$\vec{A} = 2 \vec{q}_{1}, 0$ and the sole conditions are encapsulated in
the last of equations(\ref{ortho}) and in equation (\ref{norm}).

A moment's reflection reveals that this only allows for a spiral in the plane
 defined
by $\vec{U}^{1}$ and $\vec{V}^{1}$.

When $n<4$ there are no configurations  
which  satisfy $\vec{S}^{2}(\vec{x}) =1$
identically for all sites $\vec{x}$ [excusing those having
$2(\vec{q}_{i}+\vec{q}_{j})=\vec{A}$ is equal to
a reciprocal lattice vector]
that are a superposition of exactly
two modes.

For instance, a (double checkerboard state along
the $i=1~ axis$) $\otimes$  (a spiral in the $23~~ plane$) 
has pairs $(i,j)$ for which $\vec{A} = 2(\vec{q}_{i} +\vec{q}_{j})$
is a reciprocal lattice vector.

As the number of minimizing modes $\{ \vec{q}_{m} \}$ increases, some of
the conditions may degenerate into one, e.g.  if 
$(\vec{q}_{1}+ \vec{q}_{2}) = (\vec{q}_{3}-\vec{q}_{2})$
(i.e. the modes are collinear). This degeneracy is
the a second route that might allow for Ising configurations 
which are superpositions of several ``good''minimum energy 
modes $\exp(i \vec{q} \cdot \vec{x})$.

The highly degenerate Ising ground states that
we have constructed previously can fall under
either one of these categories.

If neither one of these situations occurs, Ising states
cannot be superpositions of several minimum energy modes:
we will be left with too many equations of 
constraints with too few degrees of freedom.

For three pairs of minimizing modes, none of which is half
a reciprocal lattice vector,  $\{ \pm \vec{q}_{m} \}_{m=1}^{3}$ 
with
\begin{eqnarray}
\vec{q}_{w} \pm \vec{q}_{t} \neq \vec{q}_{r} \pm \vec{q}_{s} \neq 2 \vec{q}_{w}
\end{eqnarray}
for all $w \neq t$, ~ and $r \neq s$, 
conditions similar to those that previously written for
 $\vec{A} = \vec{q}_{1} \pm \vec{q}_{2}$, now hold for all 
$(\vec{q}_{w} \pm \vec{q}_{t})$.

\begin{eqnarray}
\vec{U}_{\alpha} \cdot \vec{U}_{\beta} = \vec{U}^{2}_{\alpha}
 ~\delta_{\alpha,\beta} \nonumber
\\ \vec{V}_{\alpha} \cdot \vec{V}_{\beta} = \vec{V}^{2}_{\alpha}
 ~\delta_{\alpha,\beta} \nonumber
\\ \vec{U}_{\alpha} \cdot \vec{V}_{\beta} =0
\end{eqnarray}
The relation $\vec{U}_{\alpha} \cdot \vec{V}_{\alpha}=0$
 ($\alpha = \beta$ in the last eqn above) is 
enforced by setting $\vec{A} = 2 \vec{q}_{\alpha}$.
Thus, when exactly three pairs of minimizing wave-vectors 
satisfying the equation are present,
the vectors $\{ \vec{U}_{\alpha},\vec{V}_{\alpha} \}$ define 
a 6-dimensional space, and hence $n \ge 6$. For $p$ pairs of minimizing
wave-vectors, $n$ must be at least $2p-$dimensional.
This bound is saturated when $\vec{S}$ is (a spiral state in the $12-plane$)

$\otimes$ (a spiral in the $34-plane$) $\otimes ... \otimes$ (a spiral in the
$2p-1,2p~~ plane$), i.e. 
\begin{eqnarray}
(a_{1} \cos(\vec{q}_{1} \cdot \vec{x}+ \phi_{1}), a_{1} \sin(\vec{q}
 \cdot \vec{x} + \phi_{1}),... \nonumber
\\ ,a_{p} \cos(\vec{q}_{p} \cdot \vec{x} + \phi_{p}),
a_{p} \sin(\vec{q}_{p} \cdot \vec{x} + \phi_{p}))
\end{eqnarray}
with $\sum_{\alpha=1}^{p} a_{\alpha}^{2} =1$.
When wave-vectors with $\vec{q}_{w} \pm \vec{q}_{t} = \vec{q}_{r} \pm
\vec{q}_{s}$ or $\vec{q}_{w} \pm \vec{q}_{t} = 2 \vec{q}_{r}$ are present,
pairs of conditions degenerate into single linear 
combinations. 

So far we have assumed that for all $i$ and $j$, 
$\vec{A} = 2(\vec{q}_{i}+ \vec{q}_{j})$ is not a reciprocal lattice 
vector, s.t. $\sin(\vec{A} \cdot \vec{x})$ is not identically
zero at all $\vec{x} \in Z^{d}$. 

We term the such a $p=2$ configuration a bi-spiral. It is
simple to see by counting the number of degrees of freedom for
$n=4$, that the bi-spirals overwhelm states having only
one mode $\pm \vec{q}_{1}$. This is a  simple instance of
a general trend: High $p$ states are statistically preferred.
Moreover, as we shall see later, they are more stable
against thermal fluctuations.

{\bf{Summary and Outlook}}

We have outlined a way to 
determine all $O(n\ge 2)$ 
ground states for a given kernel
$V(\vec{x},\vec{y}) = V(\vec{x}-\vec{y})$.

Whenever $n \ge 2$, any ground state configuration can be decomposed
into Fourier components,
${\vec S}^{g}({\vec x})= \sum_{i=1}^{|{\cal{M}}|} \left \{ 
\cos[{\vec q}_{i} \cdot \vec x] + {\vec b}_i\sin[{\vec q}_{i}
\cdot \vec x]
\right \}$

${\vec q}_{i}$ are chosen from the set of wave vectors
which minimize $v(\vec k)$.

$|{\cal{M}}|$ is the number of minimizing modes 
(the ``measure'' of the modes on the minimizing
surface $M$.  

So long as these wave-vectors ${\vec q}_{i}$
which minimize $v(\vec k)$ 
are ``non-degenerate'', in the sense
that the sum of any pair of wave vectors,  
$\vec{q}_{i} \pm \vec{q}_{j}$  is not equal to
the sum of any other pair of wave vectors, and ``incommensurate'' in 
the sense that for all $i$ and $j$, 
$2(\vec{q}_{i}+\vec{q}_{j})$ is not equal to
a reciprocal lattice vector, then the 
that the condition $[\vec{S}^{g}(\vec{x})]^{2}=1$
can be satisfied only if $|{\cal{M}}| \le n/2$.  (In our toy model 
of the doped Mott insulator, these conditions 
are always satisfied for $Q < 4$.)
Thus, for $n\le 3$ 
only simple spiral ($|{\cal{M}}|=1$) ground-states are permitted, while for
$n=4$, a double spiral saturates the bound.
Thus, generically, for $2 \le n<4$ all
ground states will be spirals
containing only one mode. 

The reader  should bear in mind that 
in the usual short range ferromagnetic
case,  the ground states
are globally  $SO(n)$ symmetric and are
labeled by only $(n-1)$ continuous
parameters.

Here, for each minimizing mode there
are $(2n-3)$ continuous internal 
degrees of freedom labeling 
all possible spiral ground states.
For $n>2$  this guarantees a much
higher degeneracy than that
of the usual ferromagnetic
ground state.

If there are many minimizing modes 
(e.g. if the minimizing manifold $M$
were endowed with $SO(d-1)$ symmetry)
then the ground state degeneracy
is even larger!

When $n\ge 4$,  there are (generically)
even many more ground states
(poly-spirals). These poly-spiral states have a degeneracies
larger than those of simple 
spiral. Their degeneracy
\begin{eqnarray}
g= p(2n-2p-1)|M|^{p},
\end{eqnarray}
where $|M|$ is the number
of minimizing modes.

\bigskip

To capitulate: we have just proved
that if frustrating interactions
cause the ground states to be modulated
then the associated  ground state
degeneracy (for $n>2$) is much
larger by comparison to the 
usual ferromagnetic ground states.

\section{Spin Stiffness} 
\label{s-stiffness}

When $Q>0$ the minimizing
modes lie of $v_{continuum}(\vec{k})$ 
lie on the surface of a sphere
 $\{ M_{Q}: \vec{q}^{2} = \sqrt{Q}\}$. 

As $Q \rightarrow 0$, this surface $M_{Q}$ 
shrinks and shrinks yet is still a $(d-1)$ 
dimensional surface of a sphere. When $Q=0$, 
the minimizing manifold evaporates into
a single point $\vec{q} =0$. This sudden change 
in the dimensionality has profound consequences. 
As we shall see shortly, it lends itself to suggest
(quite strongly) that order is inhibited for a
Heisenberg ($n=3$) realization of our model.

Before doing so, let
us indeed convince ourselves,
on an intuitive level, that the large
degeneracy in $\vec{k}-$ space
brought about by the frustration
gives rise to a reduced spin stiffness.

\subsection{Longitudinal}
Let us assume twisted boundary conditions 
\begin{eqnarray}
\vec{S}(\vec{x}) = \cos( \frac{2 \pi}{L} x + q x) \hat{e}_{1} + 
\sin(\frac{2 \pi}{L} x + qx ) \hat{e}_{2} + \delta \vec{S} \nonumber 
\\ = \hat{e}_{1} \cos(\vec{k} \cdot \vec{x}) + \hat{e}_{2} \sin (\vec{k} \cdot \vec{x}) + \delta \vec{S}
\end{eqnarray}   
with $\vec{k} = (\frac{2 \pi}{L}+ q ) \hat{e}_{1}$.
The energy cost of this state relative to the ground state
is
\begin{eqnarray}
\Delta H[\{ \vec{S}(\vec{x}) \} ] = \frac{1}{2N} \sum_{\vec{k}^{\prime}} [v(\vec{k}^{\prime})-v(\vec{q})] |\vec{S}(\vec{k}^{\prime})|^{2}
\end{eqnarray}
Ignoring $\delta \vec{S}$ contributions:
\begin{eqnarray}
\Delta H = \frac{N}{2}[v(\vec{k})-v(\vec{q})] = \frac{N}{2 \sqrt{Q}} [( \frac{2 \pi}{L} + q)^{2} - q^{2}]^{2} \approx \frac{8 \pi^{2}N}{L^{2}}. 
\end{eqnarray}
For the usual nearest neighbor ferromagnetic $XY$  model:
\begin{equation}
\vec{S}(\vec{x}) = \cos [ \frac{2 \pi x}{L}] \hat{e}_{1} + 
\sin [\frac{2 \pi x}{L}] \hat{e}_{2}.
\end{equation}
Here 
\begin{eqnarray}
v(\vec{k}) = 2 \sum_{l=1}^{3} (1- \cos k_{l}) \nonumber
\\ \Delta H_{XY} = [1-\cos(\frac{2 \pi}{L})]N \rightarrow
\frac{2 \pi^{2}N}{L^{2}} 
\end{eqnarray}
(exactly the same).

\subsection{Transverse}

\begin{eqnarray}
\vec{S}(\vec{x}) = \cos[\frac{2 \pi x}{L} + qy] \hat{e}_{1} + \sin[\frac{2 \pi x}{L} + qy] \hat{e}_{2} + \delta \vec{S}
\end{eqnarray}
\begin{eqnarray}
\vec{k} = \frac{2 \pi}{L} \hat{e}_{1} + q \hat{e}_{2}.
\end{eqnarray}
\begin{eqnarray}
v(\vec{k}) - v(\vec{q}) = \frac{16 \pi^{4}}{L^{4} \sqrt{Q}}
\end{eqnarray}
Ignoring $\delta \vec{S}$ contributions:
\begin{eqnarray}
\Delta H = \frac{8 \pi^{4} N}{L^{4} \sqrt{Q}}
\end{eqnarray}
$N \sim L^{d}$. 
\begin{eqnarray}
\Delta H \rightarrow 0
\end{eqnarray}
as $L \rightarrow \infty$ in $d=3$ [a complete loss of stiffness
against transverse fluctuations]. For a one dimensional chain 
of nearest neighbor $XY$ spins:
\begin{eqnarray}
\Delta H \sim \frac{2 \pi^{2}}{L^{2}} N = \frac{2 \pi^{2}}{L} = {\cal{O}}(\frac{1}{L}).
\end{eqnarray}
For our frustrated three-dimensional system, the transverse fluctuations obey
\begin{eqnarray}
\Delta H \sim \frac{8 \pi^{4}}{\sqrt{Q}L} = {\cal{O}}(\frac{1}{L}).
\end{eqnarray}

\subsection{ }

In principle, one may envision other impositions of twisted boundary conditions 
[say in the 3-4 plane]:
\begin{eqnarray}
\vec{S}(\vec{x}) = [\cos(qx) \hat{e}_{1} + \sin(qx) \hat{e}_{2}] \sqrt{1 -\epsilon^{2}} \nonumber 
\\+ [\cos(\frac{2 \pi y}{L}) \hat{e}_{3} + \sin(\frac{2 \pi y}{L}) \hat{e}_{4}] \epsilon + \delta \vec{S}
\end{eqnarray}
\begin{eqnarray}
\Delta H = \frac{1}{2 N} [ v(\vec{k}= \frac{2 \pi}{L} \hat{e}_{y}) - v(\vec{q})] N^{2} \epsilon^{2} \nonumber 
\\ = \frac{N}{2} \epsilon ^{2}[v(\vec{k}=0)-v(\vec{q})] = \infty
\end{eqnarray}
where in the last line we have taken the limit $L \rightarrow \infty$.
The system exhibits infinite spin stiffness to these sorts of fluctuations. 

\subsection{}
To summarize, the system responds to fluctuations with an 
effective kernel
\begin{eqnarray}
E_{low}(\vec{\delta}) 
\sim A_{\perp}~\delta_{\perp}^{4} + A_{||} ~\delta_{||}^{2}
\label{twist}
\end{eqnarray}
where $\delta_{\perp}$ and $\delta_{||}$ denote the the transverse and 
longitudinal fluctuations. 

For a nearest neighbor one dimensional system embedded in d-dimensions,
$v(\vec{k}) - v(\vec{q}) \sim A_{||} \delta_{||}^{2}$ ~~~$(A_{\perp} =0)$, 
and thus the fluctuations are even larger than in our case.

\section{Thermal fluctuations of an XY model}
\label{XY-fluct}

Let me begin by treating the ``soft-spin'' version of the
XY model, 
in which we include the non-linear interaction
\begin{eqnarray}
H_{soft} = H_{0} +  u \sum_{{\vec  x}} [\vec{S}^{2}({\vec  x})-1]^{2}
\nonumber
\\ \equiv H_{0} + H_{1}
\end{eqnarray}
with $u>0$
and forget about the normalization
conditions $|\vec{S}({\vec  x})|=1$
at all lattice sites $\vec{x}$.

(The normalized ``hard-spin'' version can be viewed as the 
$u\rightarrow \infty$ limit of the soft-spin model.)

As we have seen previously, the only generic ground 
states (for both hard- and soft-spin models)
when the spins $\vec{S}(\vec{x})$ have 
two (and also three) components are spirals

\begin{equation}
S_{1}^{ground-state}(\vec{x}) = \cos(\vec{q} \cdot \vec{x}); 
~ S_{2}^{ground-state}(\vec{x}) = \sin(\vec{q} \cdot \vec{x}). 
\end{equation}

We will expand $H_{soft}$ about these
ground states, keeping only the
lowest order (quadratic) terms
in the fluctuations $\delta S$.
The quadratic term in $\{ \delta S_{i}(\vec{k}) \}$ stemming
from $H_{soft}$
is the bilinear $\frac{u}{N} (\delta S)^{+} M (\delta S)$
where 
\begin{eqnarray}
(\delta S)^{+} = (\delta S_{1}(-\vec{k}_{1}), \delta S_{2}(-\vec{k}_{1}), 
\delta S_{1}(-\vec{k}_{2}) \delta S_{2}(-\vec{k}_{2}), \nonumber
\\
\delta S_{1}(-\vec{k}_{3}), \delta S_{2}(-\vec{k}_{3}), ..., 
\delta S_{1}(-\vec{k}_{N}), \delta S_{2}(-\vec{k}_{N}))
\end{eqnarray}
and the matrix $M$ 
reads 
$$
\pmatrix{ 4 & 0 & . & .& 1 & i & . & . & . & . \cr
 0 & 4 & . & . & i &-1 & . & . & . & .\cr 
. & . & . & . & . & . & . & .  & . & . \cr
. & . & . & . & . & . & . & . & . & . \cr
 1 & -i & . & . & 4 & 0 & . & . & 1 & i \cr
 -i & -1 & . & . & 0 & 4 & . & . & i & -1 \cr
 . & . & . & . & . & . & . & . & . & . \cr
 . & . & . & . & . & . & . & . & . & . \cr
. & . & . & . & 1 & -i & . & . & 4 & 0 \cr
 . & . & . & . & -i &-1 & . & . & 0 & 4}.
$$
A few off-diagonal sub-matrices have been omitted.
[There are two along each horizontal row (by periodic b.c.).]
The sub-matrices are $(2 \times 2)$ matrices in the internal spin indices.
The off diagonal blocks are separated from the diagonal ones
by wave-vectors $(\pm 2 \vec{q})$. 
  
Note that $\langle \vec{k}|M|\vec{k}^{\prime} \rangle = 
M(\vec{k} - \vec{k}^{\prime})$.  
Making a unitary (symmetric Fourier) 
transformation to the real space basis:
\begin{equation}
|\vec{x} \rangle \equiv N^{-1/2} \sum_{\vec{k}} e^{i \vec{k} \cdot
\vec{x}} |\vec{k} \rangle
\end{equation}
the matrix $M$ becomes block diagonal

\begin{equation}
\langle \vec{x}| M|\vec{x}^{\prime} \rangle = \hat{M}(\vec{x})
\delta_{\vec{x}, \vec{x}^{\prime}}  .
\end{equation}

Diagonalizing in the internal spin basis:
\begin{equation}
\lambda_{\pm} = 6,2.
\label{sep}
\end{equation}

These eigenvalues may be regarded, in the usual
ferromagnetic case ($\vec{q} = 0$) as 
a two step (state) potential barrier separating 
the two polarizations. I.e., the normalization 
constraint of the XY spins (embodied in $M$) 
gives rise to an effective binding
interaction. As we shall later see, 
when the number of spin components $n$
is odd, one spin component will
remain unpaired. $H_{1}$ literally 
``couple''s the spin polarizations.

Employing Eqn.(\ref{sep}), we note that the corrected 
fluctuation  spectrum $\{ \psi_{m} \}_{m=1}^{N}$
(to quadratic order)
satisfies a Dirac like equation
\begin{eqnarray}
[U^{+}~v(-i \partial_{x})~U + 2 u ~ \pmatrix{ 2 & 0 \cr
0 & 6}~ ]~ U^{+}|\psi_{m}(\vec{x}) \rangle \nonumber
\\ = E_{m}~ U^{+}|\psi_{m}(\vec{x}) \rangle,
\end{eqnarray}

Here 

\begin{eqnarray}
U = \pmatrix{ \frac{sin (2 \vec{q} \cdot \vec{x})}{2
 \cos(\vec{q} \cdot \vec{x})} & \frac{1}{2} \cr
\frac{-1 - \cos(2 \vec{q} \cdot \vec{x})}{2 \cos(\vec{q} \cdot \vec{x})} & 
\frac{1-\cos(2 \vec{q} \cdot \vec{x})}{2 \sin (\vec{q} \cdot \vec{x})}}.
\end{eqnarray}

Alternatively, expanding in the 
fluctuations $\delta S(\vec{k})$.
leads to bilinear 
$(\delta S)^{+} {\cal{H}} (\delta S)$
where   
\begin{eqnarray}
{\cal{H}}_{\vec{k},\vec{k}}= \pmatrix{v(\vec{k})+ 8u & 0\cr
0 & v(\vec{k})+ 8u }
\end{eqnarray}
along the diagonal, and  
\begin{eqnarray}
{\cal{H}}_{\vec{k},\vec{k}+2\vec{q}}= \pmatrix{ 2 u & 2 i u \cr
2 i u & -2 u}; ~ {\cal{H}}_{\vec{k},\vec{k}-2\vec{q}} = \pmatrix{ 2 u & -2 i u
\cr -2 i u & -2 u}
\end{eqnarray}
off the diagonal.
\begin{eqnarray}
{\cal{H}}_{\vec{k},\vec{k} \pm 2\vec{q}} = 2u~(\sigma_{3} \pm i \sigma_{1}).
\end{eqnarray}
\begin{eqnarray}
\exp(i \frac{\pi}{4} ~ \sigma_{1}) {\cal{H}}_{\vec{k},\vec{k} \pm 2 \vec{q}}
\exp(- i \frac{\pi}{4} ~ \sigma_{1}) \nonumber
\\ = 2u \sigma^{\pm} = 4u 
~~[~~\pmatrix{ 0&1 \cr
0&0}, \pmatrix{0& 0\cr
1&0}~~] \nonumber \\
{\cal{H}}_{\vec{k},\vec{k}} = [v(\vec{k}) + 8u] \pmatrix{ 1 & 0 \cr 
0 & 1}
\end{eqnarray}

If $Q=16$, $\vec{q} = (\pi,0,0) \in M_{Q}$, and 
$2 \vec{q} \equiv 0 (mod ~2 \pi)$. The fluctuation
matrix is diagonal in the $\vec{k}$ basis,
and the fluctuations are divergent at finite 
temperatures. An identical situation occurs for 
$\vec{q} = (\pi,\pi,0) \in M_{Q=64}$.
When $Q= 4$, $\vec{q} = (\frac{\pi}{2},0,0) \in M_{Q} $,
and it is easy to show that determining the eigenvalue spectrum
degenerates into a problem in two parameters ($\Delta_{2}, k_{1}$)
where  $\Delta_{2} \equiv  2\sum_{l=2}^{3} (\frac{3}{2}-\cos k_{l})$ 
( s.t. $\Delta(\vec{k} + \vec{q}) = \Delta_{2} +
 2 \cos k_{1}, ~\Delta(\vec{k}) = \Delta_{2} -2 \cos k_{1}$~),
The fluctuation integral about the chosen ground state exhibits a
($d-2$) dimensional minimizing manifold (parameterized, in our case,
by ($\Delta_{2},k_{1}$)~). 
Note that, at higher order commensurabilities,
the dimensionality of the 
minimizing manifold is low. In fact,
the interaction will no longer be diagonalized in $\vec{k}-$space.  

Till now, all that was stated, held for arbitrarily
large $u$- our only error was neglecting 
$O((\delta S)^{3})$ terms by comparison 
to $O((\delta S)^{2})$. Note that the main difficulty 
with the approach taken till now was the coupling
between $\vec{k}$ and $\vec{k} \pm 2 \vec{q}$:
i.e. $\vec{k}$ is coupled to 
$\vec{k} \pm 2 \vec{q}$, while $\vec{k}+ 2 \vec{q}$
is coupled to $\vec{k} + 4\vec{q}$ and $\vec{k}$, and so on. 
Unless $\vec{q}$ is of low commensurability
an exact solution to this problem is impossible. 

To make progress let us assume that $u$ is very small, 
In this case the lowest eigenstates of the fluctuation
matrix will contain only a superposition of the low lying $\vec{k}$- states
[i.e. those close to the ($d-1)$ dimensional $M$ ($Q>0$)]. 
If $\vec{k}_{1} = \vec{q}+ \vec{\delta}$ is close to $M$, 
then the only important modes in the sequence 
$\{S_{i}(\vec{k} =  \vec{k}_{1}+ 2n\vec{q})\}$ are $\vec{k}_{1}$, and
$\vec{k}_{2}=\vec{k}_{1} - 2 \vec{q} = -\vec{q}+\vec{\delta}$. 
The sub-matrix in the 
relevant sector reads
\begin{eqnarray}
\pmatrix{ v(\vec{k}_{1})+8u & 0 & 0 & 4u \cr 
0 & v(\vec{k}_{1})+8u & 0 & 0 \cr
0 & 0 & v(\vec{k}_{2})+8u & 0 \cr
4u & 0 & 0 & 0 & v(\vec{k}_{2})+8u}.
\nonumber
\end{eqnarray}
The lowest eigenvalue reads
\begin{eqnarray}
E_{low} = \frac{1}{2}[v(\vec{k}_{1})+v(\vec{k}_{2})]+4u \nonumber
\\ -\frac{1}{2}
 \sqrt{[v(\vec{k}_{1})-v(\vec{k}_{2})]^{2}+64 u^{2}}
\label{spec}
\end{eqnarray}

Equivalently, this can be determined from the direct computation
of the determinant to $O(u^{2})$: to obtain $O(u^{2})$ contributions
we need to swerve off the diagonal twice. 
\begin{equation}
\det {\cal{H}} = \sum_{i_{1},i_{2},...,i_{2N}}  \epsilon_{i_{1}i_{2}i_{3}...i_{2N}} {\cal{H}}_{1, i_{1}} {\cal{H}}_{2,i_{2}} ... {\cal{H}}_{2N, i_{2N}}.
\end{equation}
\begin{eqnarray*}
\! \! \! \! \lefteqn{\! \! \! \! \! \! \! \!
\det{\cal{H}} = \Pi_{i=1}^{N} [v(\vec{k}_{i})+8u]^{2} 
-(4u)^{2} \sum_{j}~ [v(\vec{k}_{j})+ 8u] } \\
& &  \times[v(\vec{k}_{j}+2 \vec{q})+8u]~ \Pi_{\vec{k}_{i}
 \neq \vec{k}_{j},\vec{k}_{j} + 2 \vec{q}}
[v(\vec{k}_{i})+8u]^{2}  \\
\end{eqnarray*}
The fluctuation spectrum is trivially determined
by replacing $v(\vec{k})$ by $[v(\vec{k}) - E]$ in $\det{\cal{H}}$
and setting it to zero. 

To this order we re-derive $E_{low}$. 
To higher order
\begin{eqnarray*}
\ldots  + (4u)^{4} \sum_{\vec{k}_{j_{1}} \neq \vec{k}_{j_{2}}, \vec{k}_{j_{1}} \pm
 2 \vec{q}}  [v(\vec{k}_{j_{1}})+8u]
[v(\vec{k}_{j_{1}}+ 2 \vec{q})+8u] \nonumber
\\ &&  \! \! \! \! \! \! \! \! \! \! \! \! \! \! \! \! \! \! \!
\! \! \! \! \! \! \! \! \! \! \! \! \! \! \! \! \! \! \! \! \! \! \! \! \! \!
\! \! \! \! \! \! \! \! \! \! \! \! \! \! \! \! \! \! \! \! \! \! \! \! \! \!
\! \! \! \! \! \! \![v(\vec{k}_{j_{2}})+8u] 
[v(\vec{k}_{j_{2}}+2\vec{q}) +8u] \nonumber
\\  &&  \! \! \! \! \! \! \! \! \! \! \! \! \! \! \! \! \! \! \!
\! \! \! \! \! \! \! \! \! \! \! \! \! \! \! \! \! \! \! \! \! \! \! \! \! \!
\! \! \! \! \! \! \! \! \! \! \! \! \! \! \! \! \! \! \! \! \! \! \! \! \! \!
\! \! \! \! \! \! \!
\! \! \! \! \! \! \! \! \! \! \! \! \! \! \! \! \! \! \! \! \! \! \! \!
\! \! \! \! \! \! \! \! \! \! \! \! \! \! \! \! \! \! \! \! \! \! \! \! \! \!
\! \!
\times  \Pi_{\vec{k}_{i} \neq \vec{k}_{j_{1}},\vec{k}_{j_{1}}+2\vec{q},
\vec{k}_{j_{2}},\vec{k}_{j_{2}}+
2\vec{q}} [v(\vec{k}_{i})+8u]^{2} \nonumber
\\ 
+ (4u)^{4} \sum_{j_{1}} [v(\vec{k}_{j_{1}}) + 8u]
  [v(\vec{k}_{j_{1}}+ 
 4 \vec{q})+8u] \nonumber
\\ \times \Pi_{\vec{k}_{i} \neq \vec{k}_{j},\vec{k}_{j} +
 2 \vec{q},\vec{k}_{j} 
+ 4 \vec{q}} [v(\vec{k}_{i})+8u]^{2} - \ldots
\end{eqnarray*}
(The partition function is trivially $Z = const ~[\det {\cal{H}}]^{-1/2}$.)

For any $u$, no matter how small, there exists a neighborhood
of wave-vectors $\vec{k}$ near $\vec{q}$ 
such that $|v(\vec{k})-v(\vec{k}+ 2 \vec{q})| \ll u$
and as before we may re-expand the characteristic equation
for these low lying modes, solve a simple 
quadratic equation, expand in the components
of $(\vec{k}-\vec{q})$ and obtain a simple dispersion
relation. 

If $\vec{\delta} = \vec{\delta}_{\perp} + \delta_{||}~ \hat{e}_{||}$,
with $\perp~,~||$ denoting directions orthogonal, and parallel 
to $\hat{n} \perp M$, 
then 
\begin{eqnarray}
E_{low} = A_{\perp}~ \delta_{\perp}^{4}+ A_{||}~ \delta_{||}^{2}
\label{twist*}
\end{eqnarray}
where
\begin{eqnarray}
A_{||} = \frac{1}{2} \frac{d^{2}v(|\vec{k}|)}{dk^{2}}|_{|k|=q};
 ~A_{\perp}= \frac{A_{||}}{4q^{2}}.
\end{eqnarray}

We have already found this form when examining
spin stiffness of the ``hard'' XY model.
(In both cases there is also an identical
 $\delta_{||}^{2} \delta_{\perp}^{2}$
term which may be omitted.) 
Having reassuringly re-derived this dispersion 
from another vista, let us note this dispersion is 
akin to the fluctuation
spectrum of the smectic liquid crystals 
\cite{Nielsen} which is well known to give rise to algebraic decay of correlations
at low temperatures.
In our case, one notes that $\vec{q} \neq 0$ and 
thus the correlations
should have an oscillatory prefactor. 
To be more precise, the correlator, in cylindrical 
coordinates, 
\begin{eqnarray}
G(\vec{x}) \sim \frac{4 d^{2}}{x_{\perp}^{2}}~ \exp[- 2 \eta \gamma -
\eta E_{1}(\frac{x_{\perp}^{2} Q^{1/4}}{2 x_{||}})]~\times ~ \cos[qx_{||}]
\end{eqnarray}
where $\eta= \frac{k_{B}T}{16 \pi}Q^{1/4}$, $d= \frac{2 \pi}{\Lambda}$ 
where $\Lambda$ is the  
ultra violet momentum cutoff, $\gamma$ is Euler's const., and 
\begin{eqnarray}
E_{1}(z) = -\gamma - \ln z - \sum_{n=1}^{\infty} (-)^{n} \frac{z^{n}}{n(n!)}
\end{eqnarray}
is the exponential integral.
For the uninitiated reader, 
we present this standard derivation in
the appendix. 

The thermal fluctuations $\int d^{d}k / E_{low}(\vec{k})$ 
diverge as $[- \ln |\epsilon|]$ with a
$\epsilon$ a lower cutoff on $|\vec{k}-\vec{q}|$ \cite{dive}.
Such a logarithmic divergence is also
encountered in ferromagnetic two dimensional 
$O(2)$ system if it were exposed to
the same analysis. Indeed, both 
models share similar characteristics
albeit having different physical 
dimensionality.

{\bf{summary and outlook}}

We have found that if $H_{soft}$ is indeed soft ($u \ll 1$)
then, within our continuum XY systems,
only quasi long range (algebraic) 
order will be observed at low temperatures 
in the presence of generic long range (e.g. $v_{Q}(\vec{k})$) frustration
These systems essentially display smectic phases.

\bigskip

\bigskip

{\bf{Some smectic trivia}}

Smectic A ordering involves a displacement $u$ of the smectic along
the z-direction (the direction of the molecular axis). 
It is customary to denote this displacement 
by $u(\vec{x})$.
As we shortly see, the kernel for the smectic is
identical to ours.
 
When the wave-vector $\vec{k}$ is along the z-direction
the displacement is longitudinal, and the energy is of the 
elastic form  $\frac{1}{2}B k_{||}^{2} |u(\vec{k})|^{2}$,
where $B$ is the compressibility for the smectic layers. 
When $\vec{k}$ is normal to z, the displacement is
transverse and the layer separation. No second order in
$\vec{k}_{\perp}$ the displacement costs no energy. In this
case the restoring force is associated with a director splay 
distortion. In this case, the elastic energy is 
density is  $\frac{K}{2} (\vec{\nabla} \cdot \vec{n})^{2}$
with $K$ the splay constant.. As $\hat{n}$ is normal to 
the layers, one has $\delta \vec{n} = - \vec{\nabla}_{\perp}u$
and an elastic energy $\frac{1}{2} K \vec{k}_{\perp} |u(\vec{k})|^{2}$.
Thus the kernel
\begin{eqnarray}
v(\vec{k}) = B k_{||}^{2} + K \vec{k}_{\perp}^{4}.
\end{eqnarray} 
As seen in the form
that we obtained for
$G(\vec{x})$,
the penetration depth $\lambda = \sqrt{K/B}$ determines the decay 
of an undulation distortion (splay director distortion)
imposed at the surface of the smectic.

\bigskip

\section{A Generalized
Mermin-Wagner-Coleman Theorem}
\label{M-W-low}

We will now  generalize the 
Mermin-Wagner-Coleman theorem 
\cite{Mermin-Wagner}, \cite{Coleman}:

{\bf All} systems with translationally invariant 
two-spin interactions in two dimensions
with a real twice differentiable Fourier transformed
kernel $v(\vec{k})$ show no spontaneous symmetry 
breaking at finite temperatures $T>0$.

Our approach is the 
standard one.  We will merely
keep it more general
instead of specializing
to ferromagnetic order 
or interactions
of one special sort
or another.

Following the previous
section, the magnetic field 
\begin{eqnarray}
\vec{h}(\vec{x}) = h \cos(\vec{q} \cdot \vec{x}) \hat{e}_{\alpha}
\end{eqnarray}
applied by itself would cause the spins
to take on their ground state values.

If $n=\alpha=2$ the unique spiral ground state ($\vec{S}^{g}(\vec{x})$), 
to which a low temperature  system would 
collapse to under the influence
of such a perturbation is
\begin{eqnarray}
S_{1}^{g}(\vec{x}) = \sin (\vec{q} \cdot \vec{x}).
\end{eqnarray}
When $n=3$ the ground state is not unique:
\begin{eqnarray}
S_{i<n}^{g}(\vec{x}) = r_{i} \sin(\vec{q} \cdot \vec{x}) \nonumber
\\ \sum_{i=1}^{n-1} r_{i}^{2}=1
\end{eqnarray}
and a magnetic field may be applied along two directions,
with all the ensuing steps trivially modified. 

With the magnetic field applied 
\begin{eqnarray}
H = \frac{1}{2} \sum_{\vec{x},\vec{y}} \sum_{i=1}^{n}
V(\vec{x}-\vec{y}) S_{i}(\vec{x}) S_{i}(\vec{y}) - 
\sum_{\vec{x}} h_{n}(\vec{x}) S_{n}(\vec{x}).
\end{eqnarray}

Note that the knowledge of the ground state
is not imperative in providing the forthcoming
proof \cite{smart}.

The standard idea \cite{ID} that we are about to exploit is the 
rotational invariance of the measure.

\begin{eqnarray}
\int d \mu~ ~\cdot    = ~ Z^{-1}  \int \Pi_{\vec{x}} d^{n}S(\vec{x})
\delta(S^{2}(\vec{x})-1) e^{-\beta H}
\end{eqnarray}
The generators of rotation in the $[\alpha \beta]$ plane
are 
\begin{eqnarray}
L_{\alpha \beta} \equiv S_{\alpha} \frac{\partial}{\partial S^{\beta}}
-S_{\beta} \frac{\partial}{\partial S^{\alpha}}.
\end{eqnarray}
\begin{eqnarray}
0= \frac{d}{d \theta} \int d^{n}S ~\delta(\vec{S}^{2}-1)~
\nonumber
\\
f(S_{1},...,S_{\alpha} \cos \theta + S_{\beta} \sin \theta,...
\nonumber
\\, S_{\beta} \cos \theta - S_{\alpha} \sin \alpha,...,S_{n}).
\end{eqnarray}
\begin{eqnarray}
0= \int d^{n}S \delta (\vec{S}^{2}-1) L_{\alpha \beta} f(\vec{S}).
\end{eqnarray}
Now let us consider in particular the 
generators of rotation from the axis
of the applied field to 
another internal spin axis
\begin{eqnarray}
(\vec{L}_{\vec{x}})_{\alpha} = S_{\alpha}(\vec{x}) \frac{\partial}
{\partial S_{\beta}(\vec{x})} - S_{\beta}(\vec{x})
 \frac{\partial}{\partial S_{\alpha}(\vec{x})}.
\end{eqnarray}
In the up and coming $\perp$ will denote the
the projection 
along the $\beta$ direction.

Let us define the following 
operators
\begin{eqnarray}
\vec{A}(\vec{k}) \equiv \sum_{\vec{x}} \exp[i \vec{k} \cdot \vec{x}]
\vec{S}_{\perp}(\vec{x}) \nonumber
\\ \vec{B}(\vec{k}) \equiv \sum_{\vec{x}} \exp[i (\vec{k} + \vec{q}) \cdot \vec{x}]
\vec{L}_{\vec{x}}(\beta H).
\end{eqnarray}

By the Schwarz inequality.
\begin{eqnarray}
| \langle \sum_{i= \alpha, \beta} A_{i}^{*}B_{i} \rangle |^{2} ~\le ~~ \langle 
\sum_{i= \alpha, \beta} A_{i}^{*}A_{i} \rangle * 
\langle \sum_{i=\alpha,\beta} B_{i}^{*}B_{i} \rangle
\end{eqnarray}
We will let $i=\alpha,\beta$ in the sum span a two element
subset of the $n$ spin components.
For any functional $C$:
\begin{eqnarray}
\vec{L}_{\vec{x}}(e^{-\beta H}C) = e^{-\beta H} \{ \vec{L}_{\vec{x}}(C)
+ C \vec{L}_{\vec{x}}(-\beta H)\}.
\end{eqnarray}
\begin{eqnarray}
0 = \int \Pi_{\vec{x}}~ d^{n}S(\vec{x})~ \delta (\vec{S}^{2}(\vec{x})-1)
~\vec{L}_{\vec{x}}[e^{-\beta H}C]
\end{eqnarray}

\begin{eqnarray}
\langle C B(\vec{p}) \rangle = \langle \sum_{\vec{x}} \exp[i \vec{p}
\cdot \vec{x}] ~\vec{L}_{\vec{x}}(C) \rangle
\end{eqnarray}
\begin{eqnarray}
\vec{L}_{\vec{y}}^{\beta}[\beta H] = 2 \times 1/2 \times
\beta \sum_{\vec{z}} \{ [S_{n}(\vec{y})S_{\beta}(\vec{z}) \nonumber
\\ - S_{\beta}(\vec{y}) S_{n}(\vec{z})] V(\vec{y}-\vec{z}) - h_{n}(\vec{x})
S_{\beta}(\vec{x})\}.
\end{eqnarray}
\begin{eqnarray}
\sum_{i= \alpha,\beta}  \langle L_{\vec{x}}^{i}(L_{\vec{y}}^{i}(\beta H)) 
\rangle~ =
 \beta \langle [\sum_{i=\alpha,\beta} S_{i}(\vec{x}) S_{i}(\vec{y})
 V(\vec{x}-\vec{y})
\nonumber
\\ - h(\vec{x}) S_{n}(\vec{x})] \rangle
\end{eqnarray}
\begin{eqnarray}
\langle \vec{B}(\vec{k})^{*} \cdot \vec{B}(\vec{k}) \rangle = \beta \sum_{\vec{x},\vec{y}} 
 \{ (\cos (\vec{k} + \vec{q}) \cdot (\vec{x}-\vec{y})-1) \nonumber
\\ \Big[ \sum_{i=\alpha, \beta} \langle S_{i}(\vec{x}) S_{i}(\vec{y}) \rangle  \Big]  V(\vec{x}-\vec{y}) \}   \nonumber
\\ -
h(\vec{x}) \langle S_{n}(\vec{x}) \rangle 
 ~~\ge~ 0
\end{eqnarray}

Henceforth, for simplicity, we specialize 
to $n=2$.

Fourier expanding the interaction
kernel 
\begin{eqnarray}
V(\vec{x}-\vec{y}) = \frac{1}{N} \sum_{\vec{t}} v(\vec{t}) e^{i \vec{t} \cdot
(\vec{x}-\vec{y})}
\end{eqnarray} 
and substituting
\begin{eqnarray}
\langle \vec{S}(\vec{x}) \cdot \vec{S}(\vec{y}) \rangle = \frac{1}{N^{2}}
\sum_{\vec{u}} \langle |\vec{S}(\vec{u})|^{2} \rangle e^{i \vec{u} \cdot 
(\vec{x} - \vec{y}) }
\end{eqnarray}
we obtain that 
\begin{eqnarray}
0 \le \langle \vec{B}(\vec{k})^{*} \cdot  \vec{B}(\vec{k}) \rangle = 
\beta \Delta_{\vec{k}}^{(2)} E = \nonumber
\\ \frac{\beta}{2N} \sum_{\vec{u}} 
 \Big[ v(\vec{u} + \vec{k}) + v(\vec{u}
- \vec{k}) \nonumber
\\ - 2 v(\vec{u}) \Big] \langle |\vec{S}(\vec{u})|^{2} \rangle
- \beta h_{\vec{q}} \langle S_{n}(-\vec{q}) \rangle 
\label{B*B}
\end{eqnarray}
where $\Delta_{\vec{k}}^{(2)} E$ measures 
the finite difference 
of the internal energy with respect to 
a boost of momentum $\vec{k}$. 
\begin{eqnarray}
\langle \vec{A}(\vec{k})^{*} \cdot \vec{A}(\vec{k}) \rangle = \sum_{\vec{x},\vec{y}}
\langle \vec{S}_{\perp}(\vec{x}) \cdot \vec{S}_{\perp}(\vec{y})
\rangle
\nonumber
\\ \times \exp[i \vec{k} \cdot (\vec{x}-\vec{y})].
\end{eqnarray}
\begin{eqnarray}
 \langle \vec{A}(\vec{k})^{*} \cdot \vec{B}(\vec{k}) \rangle= 
\langle \sum_{i,\vec{x}}
L_{\vec{x}}^{i}(\vec{S}_{\perp}(\vec{x})) \exp[i (\vec{k} + \vec{q}) 
\cdot \vec{x}] \rangle
\nonumber \\
= m_{\vec{q}}
\end{eqnarray}
where
$m_{\vec{q}} \equiv \langle S_{n}(\vec{q}) \rangle$. 
Note that with our convention
for the Fourier transformations,
a macroscopically modulated state of
wave-vector $\vec{q}$, $m_{q} ={\cal{O}}(N)$
as is the energy difference in Eqn.(\ref{B*B}).

The Schwarz inequality reads
\begin{eqnarray}
2 N |m_{\vec{q}}|^{2}  \Big( \beta  \sum_{\vec{k}} 
 ( \langle |\vec{S}(\vec{u})|^{2} \rangle (v(\vec{p}+\vec{u}) \nonumber 
\\ +v(\vec{p}-\vec{u}))-2 v(\vec{u})]+ 2 |h||m_{q}| 
\Big)^{-1}\nonumber
\\ \le N \sum_{\vec{k}} \sum_{\vec{x},\vec{y}} \langle \vec{S}_{\perp}(\vec{x})
\cdot \vec{S}_{\perp}(\vec{y}) \rangle e^{i \vec{k} \cdot (\vec{x}-\vec{y})}
\nonumber 
\\ = N \sum_{\vec{x}} \langle \vec{S}_{\perp}^{2}(\vec{x}) \rangle.
\label{ineq}
\end{eqnarray}

Explicitly, as the integral $\int^{|\vec{k}| > \delta}$ 
$\frac{d^{d}k}{(2 \pi)^{d}} ...$ is non-negative 
(as $\langle \vec{B}(\vec{k})^{*} \cdot \vec{B}(\vec{k}) \rangle \ge 0$
the denominator in Eqn.(\ref{ineq})
is always positive for each individual 
value of $\vec{k}$), and
as $\langle \vec{S}_{\perp}^{2}(\vec{x}) \rangle  \le 1$,
we obtain in the thermodynamic limit

\begin{eqnarray}
\frac{2}{\beta} |m_{\vec{q}}|^{2}  \int^{|\vec{k}| < \delta}
\frac{d^{d}k}{(2 \pi)^{d}} \Big[ \int \frac{d^{d}u}{(2 \pi)^{d}} \nonumber
\\  \langle |\vec{S}(\vec{u})|^{2} \rangle  (v(\vec{k}+\vec{u}) 
+v(\vec{k}-\vec{u})-2 v(\vec{u}))  \nonumber
\\ +2 |h| |m_{q}| \Big]^{-1}
 \le 1.
\label{main}
\end{eqnarray}
 
Taking $\delta$ to be small we may bound from above 
(for each value of $\vec{k}$) the positive
denominator in the square brackets 
and consequently
 
\begin{eqnarray}
 \int^{|\vec{k}| < \delta}
\frac{d^{d}k}{(2 \pi)^{d}} \Big[ \int \frac{d^{d}u}{(2 \pi)^{d}} \nonumber
\\  \langle |\vec{S}(\vec{u})|^{2} \rangle  (v(\vec{k}+\vec{u}) 
+v(\vec{k}-\vec{u})-2 v(\vec{u}))  \nonumber
\\ +2 |h| |m_{q}| \Big]^{-1}
\ge \int^{|\vec{k}| < \delta}
\frac{d^{d}k}{(2 \pi)^{d}} \Big[ \int \frac{d^{d}u}{(2 \pi)^{d}} \nonumber
\\ A_{1}  k^{2} \lambda_{\vec{u}} \langle |\vec{S}(\vec{u})|^{2} \rangle  
+2 |h| |m_{q}| \Big]^{-1}
\end{eqnarray}

with $\lambda_{\vec{u}}$ chosen to be the largest 
principal eigenvalue of the $d \times d$ 
matrix $\partial_{i} \partial_{j} [v(\vec{u})]$,
and $A_{1}$ a constant. 

For a twice differentiable $v(\vec{u})$, and for $|\vec{k}| \le \delta$ 
where $\delta$ is finite,
\begin{eqnarray}
(v(\vec{k}+\vec{u}) 
+v(\vec{k}-\vec{u})-2 v(\vec{u})) 
\le A_{1} \lambda_{\vec{u}} k^{2} \le B_{1} k^{2}
\end{eqnarray}
for all $\vec{u}$ within the Brillouin Zone  
with the additional positive constant $B_{1}$
introduced. 

In $d \le 2$, the integral 
\begin{eqnarray}
\int^{|\vec{k}| < \delta} \frac{d^{d}k}{B_{1}k^{2}}
\label{divergence1}
\end{eqnarray}
diverges making it possible to satisfy 
eqn.(\ref{main}) when the external magnetic
field $h \rightarrow 0$ only if 
the magnetization $m_{q}=0$.
If finite size effects 
are restored, in a system
of size $N= L \times L$ where 
the infrared cutoff in
the integral is ${\cal{O}} (\frac{2 \pi}{L}, \frac{2 \pi}{L})$
the latter integral diverges as ${\cal{O}}(\ln N)$.
This implies that the upper bound on $|m_{\vec{q}}|$
scales as ${\cal{O}}(N / \sqrt{\ln N})$, much lower than
the $ {\cal{O}}(N)$ requisite for finite 
on-site magnetization. For further 
details see \cite{smart}. 

If there are $M \ge 2$ pairs
of minimizing modes and $2p +1 \ge  n \ge 2p$
(with an integer $p$) then we may apply an infinitesimal 
symmetry breaking magnetic field along, at most, $\min\{p,M\}$ 
independent spin directions ($\alpha$). Employing the
spin rotational invariance within each  plane 
 $[\alpha \beta]$ associated with any individual 
mode $\vec{\ell}$ (au lieu of a specific $\vec{q}$)
we may produce a bound similar that 
in Eqn.(\ref{main}) wherein $\langle |\vec{S}(\vec{u})|^{2}
\rangle$ will be replaced by $\sum_{i= \alpha,\beta} 
\langle |S_{i}(\vec{u})|^{2} \rangle$ and $|m_{\vec{q}}| \rightarrow
|m_{\alpha}(\vec{\ell})|$. As mentioned earlier,
knowing the actual ground state (spiral or not) 
is not imperative \cite{smart}. 

The finite temperature behavior 
of a quantum system is, in many 
respects, similar to that of a 
classical system. One could directly tackle the
$n=3$ quantum case by applying the Bogoliubov
inequality
\begin{eqnarray}
\frac{\beta}{2} \langle \{A,A^{\dagger}\} \rangle * \langle
[~[C,H],C^{\dagger}]\rangle \ge | \langle [C, A]\rangle |^{2}
\end{eqnarray}

with $[~,~]$ and $\{~,~\}$ the commutator and anticommutator
respectively. Setting $A = S_{1}(\vec{k})$ and $B= S_{3}(\vec{k}+\vec{q})$
 we will once
again obtain Eqn(\ref{main}) 
with the classical spins replaced 
by their quantum counterparts.

\section{Mermin-Wagner-Coleman Bounds in High Dimensions}

In any dimension,  
Eqn.(\ref{main}) reads in the limit $h \to  0$ 
\begin{eqnarray}
2 |m_{\vec{q}}|^{2} T \int \frac{d^{d}k}{(2 \pi)^{d}} ~ ~ 
\frac{1}{\Delta_{\vec{k}}^{(2)}E} \le 1
\label{general**}
\end{eqnarray}
with the shorthand $\Delta_{\vec{k}}^{(2)}E$ defined by
Eqn.(\ref{B*B}).

By parity invariance and the
``real''ness of the spins 
$\{\vec{S}(\vec{x})\}$:
\begin{eqnarray}
\sum_{\vec{u}} v(\vec{k} + \vec{u}) \langle
|\vec{S}(\vec{u})|^{2} \rangle = \sum_{\vec{u}} v(\vec{k} - \vec{u})
\langle
|\vec{S}(\vec{u})|^{2} \rangle.
\end{eqnarray}

Thus the denominator of Eqn.(\ref{general**}) 
reads $\Delta_{\vec{k}}^{(2)}E = [2(E_{\vec{k}}-E_{0})]$
where $E_{\vec{k}}$ is the internal 
energy of system after udergoing a 
boost of momentum $\vec{k}$. 

In some instances when the dispersion relation 
about an assumed zero temperature ground
state is inserted into Eqn.(\ref{general**}),
we will find that the integral in Eqn.(\ref{general**})
diverges. This merely signals that at arbitrarily
low temperatures we cannot assume the zero temperature
ground state with the natural dispersion relation
$\Delta E$ for fluctuations about it.

A case in point is the dispersion relation
derived in Eqn.(\ref{spec}) for the Coulomb
Frustrated Ferromagnet. The reader might
recognize the denominator
in Eqn.(\ref{main}) as nothing
but a finite temperature 
extension of the zero 
temperature dispersion  
relations ($\Delta_{\vec{k}}^{(2)} E$) 
derived in Eqn.(\ref{twist}) and Eqn.(\ref{spec}) 
with the boost parameter $\vec{k}$ portrayed 
by $\vec{\delta}$. In general, at zero temperature,
\begin{eqnarray}
\langle |\vec{S}(\vec{k})|^{2} \rangle = \frac{N^{2}}{2}
[\delta_{\vec{k},\vec{q}}+ \delta_{\vec{k},-\vec{q}}]
\end{eqnarray}
and the integral 
in Eqn.(\ref{general**})
is nothing but 
\begin{eqnarray}
\int \frac{d^{d}k}{v(\vec{k}+\vec{q}) + v(\vec{k} - \vec{q}) - 2
v(\vec{q})}.
\label{my_integral}
\end{eqnarray}

Whenever this integral diverges, a bold assumption
of an almost ordered ground state at arbitrarily
low temperatures ($T = 0^{+}$) 
with the ensuing zero temperature 
dispersion relation 
($\Delta_{\vec{k}}^{(2)}E = [2(E_{\vec{k}}-E_{0})]$) 
about that ordered state is flawed: Eqn.(\ref{general**}) 
is strongly violated.

This poses no problem for all 
canonical $ d>2$ dimensional models
where the integral in Eqn.(\ref{my_integral})
is finite. For the three dimensional Coulomb 
Frustrated Ferromanget and other high dimensional 
models the divergence of this integral
hints possible non-trivialities.

\section{$O(n=3)$ fluctuations}
\label{n=3}

Let us now return to the more naive 
``soft-spin'' $O(3)$ models
in order to witness an intriguing 
even-odd binding-unbinding effect 
that could have otherwise been 
missed.

As we have proved, for an $n=3$ system the generic 
ground states are simple spirals.
If we rotate the helical ground-state 
to the $1-2$ plane; the single quadratic 
term in $\delta S_{i=3}(\vec{x})$ 
is $\sum_{\vec{x}} (\delta S_{i=3}(\vec{x}))^{2}$.
\begin{equation}
\lambda_{i \ge 3} = 2 =  \lambda_{-} = \lambda_{\min}.
\end{equation}

Here, all that follows holds for
arbitrarily large $u$- the only approximation that 
we are making is neglecting $O((\delta S)^{3})$
terms by comparison to quadratic terms: i.e. assuming
that $\delta S(\vec{x}) \ll 1$. 
Unlike the above treatment of the XY spins,
no small $u$ is necessary in order to 
make headway on the Heisenberg problem. 

For $n=3$ we find that the fluctuation
eigenstates of the are the products 
of an eigenstate of ${\cal{H}}$ within the plane 
of the spiral ground state  
and a fluctuation  eigenstate of $v(\vec{k})$
along the direction orthogonal
to the spiral plane. Written formally, 
to quadratic order, the fluctuation eigenstates are
\begin{equation}
|\psi_{m} \rangle  \otimes |\delta S_{3}(\vec{k}) \rangle.
\end{equation}
Fluctuations along the $i=3$ axis are orthogonal (in a geometrical and
formal sense) to the ground-state plane. This is expected 
as fluctuations in any hyper-plane perpendicular to the $[12]$ plane
do not change, to lowest order, the norm of the spin. 
$|\delta S_{3}(\vec{k}) \rangle$ is literally the ``odd'' man out.
As foretold, this is a general occurrence. Whenever the number
of spin  components is odd, one unpaired spin component 
is unaffected by the interaction enforcing 
the spin normalization constraint. 

Within the $i=3$ subspace $\langle \vec{k}| M |\vec{k}^{\prime} \rangle =
2 \delta_{\vec{k},\vec{k}^{\prime}}$ ~and our previous analysis follows.
The dispersion $E_{k}^{i=3} = [v(\vec{k})+ 2~ u ~ \lambda_{i=3}]$
does not have a higher minimum than with $E_{m}$ \cite{explain}. 

Both have the same $\lambda$ value and the $\delta S_{i=3}(\vec{x})$
fluctuation is minimized at wave-vectors $ \vec{\ell} \in M_{Q}$ s.t.
$v(\vec{\ell}) = v(\vec{q}) = \min_{\vec{k}} v(\vec{k})$. 
As $|\psi_{m} \rangle$ is a normalized superposition
of $|\delta \vec{S}(\vec{k}) \rangle$ modes
(the latter spin vectors being in the $1-2$ plane) 
and $v(\vec{k})$ is diagonal, 
and attains its minimum 
at $\vec{\ell} \in M_{Q}$:
\begin{equation}
\min_{m} \{ E_{m} \} \ge E_{\vec{\ell}\in M_{Q}}^{i=3}.
\end{equation}
Here, we invoked the trivial inequality 
\begin{eqnarray}
\min_{\psi} \langle \psi| [H_{0}+H_{1}]| \psi \rangle  \ge \min_{\phi}
\langle \phi|H_{0}|\phi \rangle + 
\min_{\xi}  \langle \xi|H_{1}|\xi \rangle. 
\end{eqnarray} 
Thus, there exist Goldstone modes corresponding to $\delta S_{i=3}$ 
fluctuations, and one
must adjust additive constants s.t.~ 
$\min_{\vec{k}}\{ E_{\vec{k}}^{i=3} \}= 0$.

The resulting fluctuation integral reads
\begin{equation}
\langle [\delta S_{i=3}(\vec{x}=0)]^{2} \rangle = 
k_{B}T \int \frac{d^{3}k}{(2 \pi)^{3}}~~ \frac{1}{v(\vec{k})
-v(\vec{q})}. 
\label{fluctuations}
\end{equation}

(where we noted translational invariance

\mbox{$ \langle \Delta S_{i}(\vec{k})
  \Delta S_{i}(\vec{k}^{\prime}) \rangle =
  \delta_{\vec{k}+\vec{k}^{\prime},0} \langle |\Delta
S_i(\vec{k})|^{2} \rangle$}
and employed equipartition). As pointed out earlier, when $Q>0$ in 
$v_{continuum}(\vec{k})$
the minimizing manifold is $(d-1)$ dimensional. 
The fluctuation integral receives divergent
contributions from the low energy modes nearby. 
By quadratic expansion about the minimum along $\hat{n} \perp M_{Q}$,
a divergent one-dimensional integral for the bounded
 $\langle (\Delta \vec{S}(\vec{x}=0))^{2} \rangle$ signals that
are quadratic fluctuation analysis calculation is
inconsistent. We are led to the conclusion that
higher order constraining terms are imperative:
We cannot throw away cubic and quartic spin fluctuation terms
$(\delta S^{3,4}(\vec{x}))$
relative to the quadratic  $(\delta S^{2}(\vec{x}))$
terms (notwithstanding the fact
that all of these terms appear with $O(u)$ prefactors 
irrespective of how large $u$ is). The spin fluctuations 
$\delta S_{i=3}(\vec{x})$ are of order unity at all finite
temperatures and  {\bf{$T_{c}(Q>0) \simeq 0$}}.

Note that this is ``almost  a theorem''. 
Here we do not demand that $u$ be small
(only $\delta S$). This is an important point.
$|\psi_{m} \rangle$ is an eigenstate for
arbitrarily large $u$.  To quadratic order
in the fluctuations,
the minimum belongs to 
$|\delta S_{3} \rangle $
(or is degenerate with it). 

The divergent fluctuations here
signal that ${\cal{O}}(\delta S^{4}) ={\cal{O}}(\delta S^{2})$.
Thus assuming that $\delta S \lesssim \sqrt{J/u}$
(with $J=1$ the exchange constant) we  reach a 
paradox. Thus, if the integral in Eqn.(\ref{fluctuations})
diverges for an $O(3)$ model then  at all finite $T$:
$\delta S \ge \sqrt{J/u}$.

When $Q=0$ the minimizing manifold 
shrinks to a point- the number of
nearby low energy modes is small 
and our fluctuation integral converges
in $d>2$. This is in accord with the
well known finite temperature phase 
transition of the nearest neighbor
Heisenberg ferromagnet: $T_{c}(Q=0) = O(1)$
(or when dimensions are fully restored-
it is of the order of the exchange constant).

Notice that a discontinuity in $T_{c}$ occurs as
$M_{Q} \rightarrow M_{Q=0} \equiv(\vec{q} = 0)$.

We anticipate that small lattice corrections 
($\lambda \neq 0$) in $v_{Q}(\vec{k})$ to 
yield insignificant modifications to $T_{c}(Q)$: 
one way to intuit this is to estimate $T_{c}$ by
the temperature at which the fluctuations,
as computed within the quadratic Hamiltonian  
 $\langle \delta \vec{S}^{2}(\vec{x}=0) \rangle = O(1)$.

{\bf{Summary and Outlook}}

 We have argued that in (essentially
hard spin) Heisenberg realizations
of our models no long range order is possible 
in the continuum limit:  $T_{c}(Q>0)=0$. 
More generally we claim that
if the integral
\begin{eqnarray}
\int \frac{d^{d}k}{v(\vec{k})-v(\vec{q})}
\end{eqnarray}
diverges then no long range order is possible at
finite temperature.
If lattice effects are mild then 
$T_{c}(Q>0)$ is expected to be small.
In the unfrustrated ($Q=0$) Heisenberg
ferromagnet in $d=3$: $T_{c} = O(1)$.

Fusing these facts, a discontinuity in $T_{c}$:
\begin{eqnarray}
 \delta T_{c} \equiv \lim_{Q \rightarrow 0}[T_{c}(0) - T_{c}(Q)]
\end{eqnarray}
is seen to exist. $T_{c}(Q=0)$ is an avoided critical
point (see fig. 1 in the introduction)).

\section{$O(n \ge 4)$ fluctuations}
\label{n=4}

The fluctuation analysis of any $O(n>2)$ system
about a spiral ground state is qualitatively similar
to that of the Heisenberg system.

For a vanishing lower cutoff $\epsilon$ on $||\vec{k}|-q|$,
the fluctuations of an $n-component$ spin ($d=3$) about a 
helical ground state are given by
\begin{eqnarray}
\frac{ \langle (\Delta \vec{S})^{2} \rangle }{k_{B}T} = \frac{(n-2) 
\sqrt{Q}}{4 \pi^{2} \epsilon} - \frac{1}{16 \pi}Q^{1/4} \ln|{\epsilon}|.
\end{eqnarray}
More generally, in $d>2$~dimensions, the leading order infrared contribution
reads
\begin{eqnarray}
\frac{(n-2)~Q^{(d-1)/4}}{2^{d}~\pi^{d/2}~\epsilon~\Gamma(d/2)} 
- \frac{\epsilon^{d-3}Q^{1/4}}{2^{d+1}~\pi^{(d-1)/2}~(d-3)~\Gamma(\frac{d-1}{2})}.
\end{eqnarray}

As noted previously, poly-spiral states
will tend to dominate at large $n$.

The reader can convince him/herself that 
for even $n$ with a $p<n/2$ poly-spiral ground state
and for  all odd $n$  the fluctuations will give 
rise to a leading order $\epsilon^{-1}$ divergence. 
The reasoning is simple: the poly-spiral states extend
along an even number of axis. If $n$
is odd then there will be at least
one internal spin direction $i$ along 
which $S_{i}^{g}=0$  and our analysis of the
Heisenberg model can be 
reproduced.

The lowest 
eigen-energy associated
with the fluctuations $|\psi_{m} \rangle$
in the $(2p)$ dimensional
space spanned by the ground state
is higher than 
the lowest eigen-energy 
for fluctuations along 
an orthogonal direction.  

$\langle \psi|H_{soft}|\psi \rangle \ge 0$.  If $|\delta S| \ll 1$,
this implies that the quadratic term in $\delta S(\vec{x})$ 
stemming from $H_{1}$ is non-negative definite.

For $S_{i}^{g}(\vec{x}) =0$, this quadratic term in 
$\delta S_{i}(\vec{x})$
is zero. 

The eigenvalue $\lambda_{\min} = \lambda_{-} =2$ 
corresponds to the zero contribution in 
$O(\delta S^{2}(\vec{x}))$ from $H_{1}$.

And once again
\begin{equation}
\min_{m} \{ E_{m} \} \ge E_{\vec{\ell}\in M_{Q}}^{i}.
\end{equation}
from the trivial inequality 
\begin{eqnarray}
\min_{\psi} \langle \psi| [H_{0}+H_{soft}]| \psi \rangle  \ge 
\nonumber
\\ \min_{\phi} \langle \phi|H_{0}|\phi \rangle  + 
\min_{\xi} \langle \xi|H_{soft}|\xi \rangle. 
\end{eqnarray}

The fluctuations of even component
spin about a $p=n/2$ poly-spiral 
ground state are more complicated.

Once again coupling between different
modes occurs. In this case they are more 
numerous.

For $n=4$, the fluctuation energy, to quadratic order,
about a bi-spiral reads

$
\! \! \! \! \! \! \! \! \! \delta H = 
\frac{1}{2N} \sum_{\vec{k}} [v(\vec{k}) - v(\vec{q})] |\delta \vec{S}(\vec{k})|^{2} 
+ \frac{u a_{1} a_{2}}{N} \{ \sum_{\vec{k}_{1},\vec{k}_{2}}[ (\delta S_{1}(\vec{k}_{1}) \delta S_{3} (\vec{k}_{2}) + \delta S_{1}(\vec{k}_{2})
\delta S_{3}(\vec{k}_{1})) \nonumber
\\ \times 
[\delta _{\vec{q}_{1}+\vec{q}_{2}+\vec{k}_{1}+\vec{k}_{2},0} + \delta_{\vec{q}_{1}+\vec{q}_{2}-\vec{k}_{1}-\vec{k}_{2},0}
 + \delta_{\vec{k}_{1}+\vec{k}_{2}+\vec{q}_{2}-\vec{q}_{1},0} +
\delta_{\vec{k}_{1} +\vec{k}_{2} +\vec{q}_{1}-\vec{q}_{2},0}] +  \nonumber
\\ + (\delta S_{2}(\vec{k}_{2}) \delta S_{4}(\vec{k}_{1})+
\delta S_{2}(\vec{k})\delta S_{4}(\vec{k}_{2}))\nonumber
\\ \times [\delta_{\vec{k}_{1}+\vec{k}_{2}+\vec{q}_{1}-\vec{q}_{2},0}+
\delta_{\vec{k}_{1}+\vec{k}_{2}+\vec{q}_{2}-\vec{q}_{1},0}-\delta_{\vec{k}_{1}+\vec{k}_{2}+\vec{q}_{1}+\vec{q}_{2},0}-\delta_{\vec{k}_{1}+\vec{k}_{2}-\vec{q}_{1}-\vec{q}_{2},0}] \nonumber
\\ -i (\delta S_{1}(\vec{k}_{2}) \delta S_{4}(\vec{k}_{1})
+ \delta S_{1}(\vec{k}_{1}) \delta S_{4}(\vec{k}_{2})) \nonumber
\\ \times  [\delta_{\vec{k}_{1}+\vec{k}_{2}+\vec{q}_{1}+\vec{q}_{2},0}-\delta_{\vec{k}_{1}+\vec{k}_{2}-\vec{q}_{1}-\vec{q}_{2},0} +\delta_{\vec{k}_{1}+\vec{k}_{2}+\vec{q}_{2}-\vec{q}_{1},0}-\delta_{\vec{k}_{1}+\vec{k}_{2}-\vec{q}_{2}+\vec{q}_{1},0}] \nonumber
\\ -i (\delta S_{2}(\vec{k}_{2}) \delta S_{3}(\vec{k}_{1})+\delta S_{2}(\vec{k}_{1}) \delta S_{3}(\vec{k}_{2}))\nonumber
\\ \times [\delta_{\vec{k}_{1}+\vec{k}_{2}
+\vec{q}_{1}+\vec{q}_{2},0}-\delta_{\vec{k}_{1}+\vec{k}_{2}-\vec{q}_{1}-\vec{q}_{2},0}+\delta_{\vec{k}_{1}+\vec{k}_{2}+\vec{q}_{1}-\vec{q}_{2},0}-\delta_{\vec{k}_{1}+\vec{k}_{2}+\vec{q}_{2}-\vec{q}_{1},0}]] \} \nonumber
\\ + \frac{2 u a_{1}^{2}}{N} \sum_{\vec{k}_{1},\vec{k}_{2}} \{  \delta S_{1}(\vec{k}_{1})\delta S_{1}(\vec{k}_{2}) [\delta_{\vec{k}_{1}+\vec{k}_{2},0}+\frac{1}{2}(\delta_{\vec{k}_{1}+\vec{k}_{2}+2 \vec{q}_{1},0}+\delta_{\vec{k}_{1}+\vec{k}_{2}-2\vec{q}_{1},0})] \nonumber
\\+ \delta S_{2}(\vec{k}_{1}) \delta S_{2} (\vec{k}_{2}) [ \delta_{\vec{k}_{1}+\vec{k}_{2},0} - \frac{1}{2}(\delta_{\vec{k}_{1}+\vec{k}_{2}+2 \vec{q}_{1},0}+\delta_{\vec{k}_{1}+\vec{k}_{2}-2 \vec{q}_{1},0})] \}
\nonumber
\\ +\frac{2 u a_{2}^{2}}{N} \sum_{\vec{k}_{1},\vec{k}_{2}} 
\{ \delta S_{3}(\vec{k}_{1}) \delta S_{3}(\vec{k}_{2}) [\delta_{\vec{k}_{1}+\vec{k}_{2},0} +\frac{1}{2} (\delta_{\vec{k}_{1}+\vec{k}_{2}+2 \vec{q}_{2},0}+\delta_{\vec{k}_{1}+\vec{k}_{2}-2\vec{q}_{2},0})] \nonumber
\\+ \delta S_{4}(\vec{k}_{1}) \delta S_{4}(\vec{k}_{2}) [\delta_{\vec{k}_{1}+\vec{k}_{2},0} -\frac{1}{2} (\delta_{\vec{k}_{1}+\vec{k}_{2}+2\vec{q}_{2},0}+\delta_{\vec{k}_{1}+\vec{k}_{2}-2\vec{q}_{2},0})]\} \nonumber
\\ -\frac{i u a_{1}^{2}}{N} \sum_{\vec{k}_{1},\vec{k}_{2}} 
[\delta S_{1}(\vec{k}_{2}) \delta S_{2}(\vec{k}_{1}) +\delta S_{1}(\vec{k}_{1})
 \delta S_{2}(\vec{k}_{2})](\delta_{\vec{k}_{1}+\vec{k}_{2}+2\vec{q}_{1},0}
-\delta_{\vec{k}_{1}+\vec{k}_{2}-2\vec{q}_{1},0}) \nonumber
\\ -\frac{i u a_{2}^{2}}{N} \sum_{\vec{k}_{1},\vec{k}_{2}} [\delta S_{3}(\vec{k}_{2})\delta S_{4}(\vec{k}_{1})+ \delta S_{3}(\vec{k}_{1}) \delta S_{4}(\vec{k}_{2})](\delta_{\vec{k}_{1}+\vec{k}_{2}+2\vec{q}_{2},0}
-\delta_{\vec{k}_{1}+\vec{k}_{2}-2 \vec{q}_{2},0}).
$

\bigskip

As before, we may obtain an equation for 
the eigenvalues by truncating the expansion 
for $\det [{\cal{H}}-E]$ at $O(u^{2})$.

Just as the dispersion relation for low lying states
could have been derived by setting $\det [{\cal{H}} -E]=0$
to $O(u^{2})$ and solving the resultant quadratic
equation for the low lying energy states:
$\vec{k} = \vec{q}+ \vec{\delta}$ with small $\vec{\delta}$. 
Here we may do the same. 

The characteristic equation
is easily read off: 
\begin{eqnarray}
0=\det[{\cal{H}}-E]= 
\Pi_{i=1}^{N} [(v(\vec{k}_{i}) - E)^{2}] - \nonumber
\\ 
4u^{2} a_{1}^{2} \sum_{j_{1}} [(v(\vec{k}_{j_{1}})+8u)
(v(\vec{k}_{j_{1}}+2 \vec{q}_{1}) +8u)]\nonumber
\\ \times \Pi_{i \neq j_{1}} [(v(\vec{k}_{i})-E]^{2}] \nonumber
\\ - 4u^{2}a_{2}^{2} \sum_{j_{2}} [(v(\vec{k}_{j_{2}})+8u)
(v(\vec{k}_{j_{2}}+2 \vec{q}_{2}) +8u)]  
\nonumber
\\ \times  \Pi_{i \neq j_{2}} [(v(\vec{k}_{i})-E]^{2}]
\end{eqnarray}
where we have shifted $E$ by a constant.
Notice that decoupling trivially occurs
- terms of the form $[v(\vec{k})-E)][v(\vec{k}^{\prime})-E]$
where the modes 
$\vec{k}-\vec{k}^{\prime} = \vec{q}_{1} 
\pm \vec{q}_{2}$, cancel.

For low lying states i.e.
for the terms containing 
$\vec{k}_{1} = \vec{q}_{1} + \vec{\delta}_{1}$
and $\vec{k}_{2} = \vec{q}_{2} + \vec{\delta}_{2}$:
\begin{eqnarray}
 E= E_{\min} + a_{1}^{2} [A_{||} \delta_{||;1}^{2} +
 A_{\perp} \delta_{\perp;1}^{4}] \nonumber
\\ 
+ a_{2}^{2} [A_{||} \delta_{||;2}^{2} + A_{\perp} \delta_{\perp;2}^{4}]
\end{eqnarray}
with $\delta_{||,\perp;m}$ parallel and 
perpendicular to the minimizing manifold $M$ 
at $\vec{q}_{m}$, 
trivially satisfies $\det[{\cal{H}}-E]=0$.


This dispersion relation agrees, once again, with the result 
derived from the spin wave stiffness analysis
\begin{eqnarray}
\Delta H = \frac{1}{2N}\sum_{\vec{k}} (v(\vec{k})- v(\vec{q}))
|\vec{S}(\vec{k})|^{2}
\end{eqnarray}
and by expansion of $\Delta H$ for different sorts of twists,
the dispersion relations of the two spiral
simply lumped together. When $d=3$, as in the 
$O(2)$ case ($p=1$)
this dispersion gives
rise (in the Gaussian approximation)
to diverging logarithmic
fluctuations: $O(| \ln ~\epsilon|)$.
Applying equipartition, the Gaussian spin
fluctuations in the $[2i-1,2i]$ plane:
\begin{eqnarray}
\Delta \vec{S}^{2}_{[2i-1,2i]}(\vec{x}=0) > \Delta S^{2}_{low~~[2i-1,2i]}(\vec{x}=0)
\nonumber \\ = k_{B} T
\int \frac{d^{3}k}{(2 \pi)^{3}}~~ \frac{1}{a_{i}^{2} [A_{||} \delta_{||;i}^{2}
+ A_{\perp} \delta_{\perp;i}^{4}]}.
\end{eqnarray}

For all odd $n$ and for all even $n$ with $p<n/2$ there
will be divergent fluctuations similar to those encountered
for the $O(3)$ model.

{We can now update and summarize our conclusions: 
We have just proved that if frustrating interactions
cause the ground states to be modulated
then the associated  ground state
degeneracy (for $n>2$) is much
larger by comparison to the 
usual ferromagnetic ground states.
For even $n$ we 
have found that, generically, 
the a three dimensional system
will not have long range order
when $M$ is two dimensional.
When $n$ is odd the system will
never show long range order
if $M$ is (d-1) or (d-2) dimensional.

If, in the continuum limit, the uniform
(ferromagnetic) state is higher
in energy than any other state
then, by rotational symmetry,
the manifold of minimizing modes in Fourier 
space is $(d-1)$ dimensional.

In reality, small symmetry breaking terms (e.g. $\lambda \neq 0$
in $v_{Q}(\vec{k})$) will 
always be present- these will favor ordering at a 
discrete set of $\{ \pm\vec{q}_{m}\}_{m=1}^{|M|}$.
If $n>2|M|$ then, irrespective of the
even/odd parity of $n$, then will be a
 
\begin{eqnarray}
\langle \Delta \vec{S}^{2}(\vec{x}=0) \rangle \ge
(n- 2 |M|)\int \frac{d^{d}k}{(2 \pi)^{d}}
\frac{k_{B}T}{v(\vec{k})-v(\vec{q})}.
\label{leftover}
\end{eqnarray}

This contribution is monotonically increasing in $n$; Within our
scheme, $T_{c}$ is finite and may be estimated by
the temperature at which the fluctuations are of
order unity. By tweaking the symmetry breaking
terms to smaller and smaller values, the 
fluctuation integral becomes larger and larger.
For instance, if take $\lambda \ll 1$ 
in $v_{Q}(\vec{k})$ then the integral
is very large and $T_{c}$ extremely low
(in can be made arbitrarily low).
Thus as the system will be cooled from 
high temperatures, it might
first undergo a Kosterlitz-Thouless 
like transition at $T_{KT}$ to an 
algebraically ordered state and
develop true long range order 
at critical temperatures $T_{c}<T_{KT}$.

\section{Large n limit}
\label{spherical}

So far we have seen that the even $n$
systems are more ``gapped'' than its
odd counterparts. There is never a 
paradox in the large $n$ (or spherical
model) limit. In this limit wherein
a single normalization
constraint is imposed
\begin{eqnarray}
\sum_{\vec{x}} S^{2}(\vec{x}) = N,
\end{eqnarray}
the effective number of spin components
$n$ is of the order of the number of 
sites in the system $N$. The span
of the system $N$ (the number
of Fourier modes allowed within the 
Brillouin zone) is always larger
than the number of minimizing modes
$\{\vec{q}_{i}\}$. 

In such a case we will be 
left with a divergence as
in equation (\ref{leftover})
due to the many unpaired
spin components.

In fact, within the spherical 
model, which is easily
solvable the fluctuation
integral exactly marks the 
value of the inverse critical
temperature
\begin{eqnarray}
\frac{1}{k_{B}T_{c}} = \int_{B.Z.} \frac{d^{d}k}{(2 \pi)^{d}} 
~ ~ \frac{1}{v(\vec{k})-v(\vec{q})}.
\end{eqnarray}

Thus, $T_{c} =0$ if the latter integral
diverges and our circle of ideas nicely 
closes on itself.

\section{$O(n \ge 2)$ Weiss Mean Field Theory
0f Any Translational Invariant Theory.}
\label{high_mft}

Let us begin by examining the situation simple spiral
states.  In this case
for $O(n \ge 2)$ when $T<T_{c}$:
\begin{equation}
\langle \vec{S}(\vec{x}) \rangle ~ = ~s ~ \vec{S}^{ground-state}(\vec{x}).
\end{equation}
For the particular case
\begin{eqnarray}
S_{1}^{ground-state}(\vec{x}) = \cos(\vec{q} \cdot \vec{x}) \nonumber
\\ S_{2}^{ground-state}(\vec{x}) = \sin(\vec{q} \cdot \vec{x}) \nonumber
\\ S_{i \ge 3}^{ground-state}(\vec{x}) =0 
\end{eqnarray}
Now only the $ \pm \vec{q}$ modes have finite weight. 
Repeating the previous steps 
\begin{eqnarray}
\sum_{\vec{y}} V(\vec{x}=0,\vec{y}) S_{2}^{ground-state}(\vec{y}) = 0 
\end{eqnarray}
\begin{equation}
| \langle \vec{S}(\vec{x}=0) \rangle | = | \langle
S_{1}(\vec{x}=0) \rangle | 
\end{equation}
Define 
\begin{eqnarray}
M[z] \equiv - \frac{d}{dz}  \ln [(2/z) ^{(n/2-1)} I_{n/2-1}(z)],
\end{eqnarray} 
with $[I_{n/2-1}(z)]$ a Bessel function.
The mean-field equation reads 
\begin{eqnarray}
|\langle S_{1}(\vec{x}=0) \rangle | = s = M[~|\sum_{\vec{y}}
V(\vec{x},\vec{y}) \langle S_{1}(\vec{y}) \rangle| ~].
\nonumber
\end{eqnarray}
The onset of the non-zero solutions is at
\begin{equation}
|\beta_{c} v(\vec{q})| = n.
\end{equation}
If $V(\vec{x}=0)=0$ (no on-site interaction), then 
\begin{equation}
\int d^{d}k~ v(\vec{k}) =0,
\end{equation}
implying that $v(\vec{q}) < 0 $
and $T_{c}>0$. Note that within the
mean field approximation,
$T_{c}$ is a continuous function
of the parameters.

Here the ground state is symmetric with respect to all sites.
The above is the exact value of $T_{c}$ 
within Weiss mean field theory for the helical ground-states.  

For poly-spirals we will get $p$ identical equations:
both sides of the self consistency equations
are multiplied by $a_{l}^{2}$ where
$a_{l}$ is the amplitude of the $l-th$ spiral in the
[$(2l-1),2l$] plane. As $v(\vec{q}_{m}) = v(\vec{q})$,
we will arrive at the same value of $T_{c}$ as for the
case of simple spirals.

\section{Extensions to arbitrary two spin interactions:
Spin Glasses, $| \vec{u} \rangle$ space $ \otimes O(n)$ 
topology etc.}
\label{brilliant}

Any real kernel $V(\vec{x},\vec{y})$ may be symmetrized 
$[V(\vec{x},\vec{y}+ V(\vec{y},\vec{x})]/2 \rightarrow
V(\vec{x},\vec{y})$ to a hermitian form. 

Consequently, by a unitary transformation,
it will become diagonal. The Fourier 
modes are the eigen-modes of $V$ when
it is translationally invariant.
We may similarly envisage extensions
to other, arbitrary, $V(\vec{x},\vec{y})$  
which will become diagonal in some
other complete orthogonal basis $|\vec{u} \rangle$:
\begin{eqnarray}
\langle \vec{u}_{i} | V | \vec{u}_{j} \rangle =  \delta_{ij}  
\langle \vec{u}_{i} | V | \vec{u}_{i} \rangle
\end{eqnarray}

Many of the statements that we have made hitherto
have a similar flavor in this more
general case.

For instance, the large $n$ fluctuation
integrals are of the same form
\begin{eqnarray}
\int \frac{d^{d}u}{(2 \pi)^{d}} ~ ~ \frac{1}{v(\vec{u}) - v_{\min}}
\end{eqnarray}
with the wave-vector $\vec{k}$ 
traded in for $\vec{u}$.

Once again, one may examine the topology 
of the minimizing manifold in $\vec{u}$
space. If the surface if $(d-1)$ dimensional
and $v(\vec{u})$ is analytic in its
environs then, for large $n$, 
$T_{c} =0$.

The topology of the ground state sector
of $O(n)$ models will once again be 
governed by a direct product of
the topology of the minimizing manifold
in $\vec{u}$ space with the 
spherical manifold of the 
$O(n)$ group. In the general
case in will be dramatically
rich.

We may similarly extend the Peierls
bounds to some infinite range interactions
also in this case by contrasting
the energy penalties in the now
diagonalizing $\vec{u}$
basis with those that occur
for short range systems in 
Fourier space
\begin{eqnarray}
\Delta E_{disordered} = \frac{1}{2N} 
\sum_{\vec{u}} |\vec{S}(\vec{u})|^{2} \langle
\langle \vec{u} | V_{dis}| \vec{u} \rangle - v_{min}(\vec{u})]
\nonumber
\\ 
\ge \frac{1}{2N} \sum_{\vec{k}} 
|\vec{S}(\vec{k})|^{2} [v_{short}(\vec{k}) - v_{short~\min}]
\end{eqnarray}
if they share the same lowest energy eigenstate of $V$.

\section{O(n) spin dynamics and simulations}
\label{dynamics}

For our  repeated general Hamiltonian
\begin{eqnarray}
H = \frac{1}{2} \sum_{\vec{x},\vec{y}} V(\vec{x},\vec{y}) 
\vec{S}(\vec{x}) \cdot \vec{S}(\vec{y})
\end{eqnarray}
the force on given spin 
\begin{eqnarray}
\vec{F}(\vec{z}) = -\frac{\partial H}{\partial \vec{S}(\vec{z})}
\nonumber
\\ = 
- \frac{1}{2} \sum_{\vec{y}} [V(\vec{z},\vec{y}) + V(\vec{y},\vec{z})]  \vec{S}(\vec{y}).
\end{eqnarray}
For $V(\vec{x},\vec{y}) = V(\vec{x}-\vec{y}) (= V(\vec{y}-\vec{x})$ or 
otherwise we may symmetrize $[V(\vec{x},\vec{y}+ V(\vec{y},\vec{x})]/2
\rightarrow V(\vec{x},\vec{y})$, as we have done repeatedly, 
without changing $H$). 
For a more general two spin kernel 
$V(\vec{x},\vec{y})$ which is not translationally
invariant the Fourier space index $\vec{k}$
should be replaced by the more general $\vec{u}$.  

\begin{eqnarray}
\frac{d^{2}\vec{S}(\vec{x})}{dt^{2}} \nonumber
\\ = - \sum_{\vec{y}} V(\vec{x}-\vec{y}) \vec{S}(\vec{y}) \frac{d^{2}
\vec{S}(\vec{k})}{dt^{2}} \nonumber
\\ = - \frac{1}{N}v(\vec{k}) \vec{S}(\vec{k}).
\end{eqnarray}
Alternatively, the last equation can be derived by
starting with the Hamiltonian expressed directly in 
Fourier space 
\begin{eqnarray}
H = \frac{1}{2N} \sum_{\vec{k}} v(\vec{k}) \vec{S}(\vec{k}) \cdot \vec{S}(-\vec{k}) \nonumber
\\ \frac{d \Pi(\vec{p},t)}{dt} \nonumber
\\ = \frac{d^{2} \vec{S}(\vec{p})}{dt^{2}} = -\frac{\partial
H}{\partial \vec{S}(\vec{p})} = -\frac{1}{2N}[v(\vec{p})+v(-\vec{p})]
\vec{S}(\vec{p}) \nonumber
\\ = - \frac{1}{N}v(\vec{p}) \vec{S}(\vec{p}),
\end{eqnarray}
where in the last equation the momentum $\Pi(\vec{x})$ conjugate to 
$\vec{S}(\vec{x})$ 
is trivially 
\begin{eqnarray}
\frac{\partial L}{\partial(d\vec{S}(\vec{x})/dt)}\nonumber
\\ = \frac{\partial (\frac{1}{2} \sum_{\vec{x}} [d
\vec{S}(\vec{x})/dt]^{2} - \frac{1}{2} \sum_{\vec{x}, \vec{y}}
V(\vec{x},\vec{y}) \vec{S}(\vec{x})\cdot \vec{S}(\vec{y}))}{\partial
(d\vec{S}(\vec{x})/dt)} \nonumber
\\ = \frac{d \vec{S}(\vec{x})}{dt},
\end{eqnarray}
and upon Fourier transforming $\Pi(\vec{p}) = (d\vec{S}(\vec{p})/{dt})$.
Let an arbitrary $A$ satisfy
\begin{eqnarray}
A > - \min_{\vec{k}} \{ v(\vec{k}) \}.
\end{eqnarray}
The equations of motion 
\begin{eqnarray}
\frac{d^{2}\vec{S}(\vec{k})}{dt^{2}} = -[A+v(\vec{k})] \vec{S}(\vec{k})\nonumber
\\ \equiv -\omega_{k}^{2} \vec{S}(\vec{k})
\end{eqnarray}
may be trivially integrated. [Adding the constant $A$ merely shifts 
$H \rightarrow H+ A/2$.]  
\begin{eqnarray}
\vec{S}_{un}(\vec{k},t) = \vec{S}(\vec{k},0) \cos \omega_{k} t + \frac{d \vec{S}(\vec{k},t)}{dt}|_{t=0}~ \times  \omega_{k}^{-1} \sin \omega_{k} t, \nonumber
\\
\vec{S}_{un}(\vec{k}, t+ \delta t) = \vec{S}(\vec{k},t) \cos
\omega_{k} \delta t \nonumber
\\ + \frac{\delta \vec{S}(\vec{k},t)}{\delta t} ~ \omega_{k}^{-1} \sin
\omega_{k} \delta t.
\nonumber
\end{eqnarray}
This suggests the following simple algorithm:

\bigskip

(i)   At time t, start off with initial values $\{\vec{S}(\vec{x}, t) \}$.

\bigskip

(ii)  Fourier transform to find $\{ \vec{S}(\vec{k},t) \}$.

\bigskip

(iii) Integrate to find the un-normalized $\{ \vec{S}_{un}(\vec{k},t+\delta t) \} $.

\bigskip

(iv)  Fourier transform back to find the un-normalized real-space spins
$\{ \vec{S}_{un}(\vec{x},t+\delta t) \} $. 

\bigskip

(v)   Normalize the spins: 
\begin{eqnarray}
\vec{S}(\vec{x},t+\delta t) = \frac{\vec{S}_{un}(\vec{x},t+\delta t )}{|\vec{S}_{un}(\vec{x},t+\delta t)|}.
\end{eqnarray}

\bigskip

(vi)  Compute $ \{ \delta \vec{S}(\vec{x}, t+\delta t) \} = \{ [\vec{S}(\vec{x},t+\delta t)-\vec{S}(\vec{x},t)] \} $.

\bigskip

(vii) Fourier transform to find $\{ \delta \vec{S}(\vec{k},t+\delta t)\}$.

\bigskip

(viii) Go back to (ii). 

\bigskip

\bigskip

Thus far we have neglected thermal effects. 
To take these into account, one could integrate these equations
with a thermal noise term augmented to the 
restoring force 

\begin{eqnarray}
\frac{d^{2}\vec{S}(\vec{k})}{dt^{2}} = - \frac{1}{N} v(\vec{k}) \vec{S}(\vec{k}) + \vec{F}_{\vec{k}}^{~noise}(T,t).
\end{eqnarray}

\bigskip

\bigskip

Expressed in this format, the execution of this algorithm for continuous
$O(n \ge 2)$ spins seems easier than that for a discrete Ising system.
Here the equations of motion may be integrated to produce arbitrarily
small updates at all sites.
  
This could, perhaps, be better than a brute force approach 
whereby the torque equations in angular variables
(the spins $\vec{S}(\vec{x})$ are automatically normalized)
are integrated whereby the eqs of motion would explicitly read

\begin{eqnarray}
\frac{d^{2}\phi(\vec{x})}{dt^{2}} = \sum_{\vec{y}} V(\vec{x}-\vec{y}) \sin [\phi(\vec{x}) - \phi(\vec{y})].
\end{eqnarray}
for an $O(2)$ system.

For the three-component spin system:
\begin{eqnarray}
\sin^{2}\theta(\vec{x})\frac{d^{2}\phi(\vec{x})}{dt^{2}} + \sin 2
\theta (\vec{x}) \frac{d \phi(\vec{x})}{dt} \frac{d
\theta(\vec{x})}{dt} \nonumber
\\ = \sum_{\vec{y}} V(\vec{x}-\vec{y}) \sin \theta(\vec{x}) 
\nonumber \times \sin \theta(\vec{y}) \sin[\phi(\vec{x})-\phi(\vec{y})]; \nonumber
\\ \frac{d^{2}\theta(\vec{x})}{dt^{2}} 
 = \frac{1}{2} \sin 2 \theta(\vec{x}) (\frac{d\phi
(\vec{x})}{dt})^{2}\nonumber
\\ +\sum_{\vec{y}} V(\vec{x}-\vec{y})\{ \sin \theta(\vec{x}) \nonumber
\\ \cos \theta(\vec{y})
- \cos \theta(\vec{x}) \sin \theta(\vec{y}) \cos[\phi(\vec{x})-\phi(\vec{y})]\}.
\end{eqnarray}

\section{Appendix}

 Here we follow the beautiful treatment of 
Als- Nielsen et al. $\cite{Nielsen}$.

Within the (hard spin) fully constrained XY model:

\begin{eqnarray}
G(\vec{x}-\vec{y}) = \langle \vec{S}(\vec{x}) \cdot \vec{S}(\vec{y})
\rangle 
= \langle \cos[\theta(\vec{x})-\theta(\vec{y})] \rangle.
\end{eqnarray}
Here $\theta(\vec{x}) =  \vec{q}\cdot \vec{x} + \Delta \theta(\vec{x})$,
i.e. $\Delta \theta$ denotes the phase fluctuations about
our spiral ground state and 
\begin{eqnarray}
G(\vec{x}-\vec{y}) = \cos(\vec{q} \cdot (\vec{x}-\vec{y}))) 
 \langle e^{i(\Delta \theta(\vec{x})- \Delta \theta(\vec{y})} \rangle.
\end{eqnarray}
In our harmonic approximation
$\{ \delta \theta(\vec{x})\}$
are random Gaussian variables
and only the first term in the 
cumulant expansion
is non-vanishing.

The correlator 
\begin{eqnarray}
G(\vec{x})= \exp [-\frac{1}{2}[ \langle |\Delta \theta(\vec{x})-\Delta 
\theta(0)|^{2}] \rangle] \nonumber
\\ =  \exp 
\left[ 
k_{B}T 
\int 
\frac{d^{d}k}
{(2 \pi)^{d}}~ 
~\frac{1-\cos \vec{q} \cdot \vec{x}}
{A_{||} \delta_{||}^{2}+ A_{\perp}
\delta_{\perp}^{4}} 
\right].
\end{eqnarray}
Now let us shift variables $\vec{k} \rightarrow  
\vec{k} - \vec{q} \equiv \delta$, and for purposes
of convergence explicitly introduce an upper 
bound on 
$k_{\perp}$: ~$0<k_{\perp}< \Lambda$ 
\begin{eqnarray} 
I(\vec{x}_{\perp},x_{||}) \equiv \int \frac{1- \cos(\vec{q} \cdot \vec{x})}
{A_{||} k_{||}^{2}+ A_{\perp} k_{\perp}^{4}}.
\end{eqnarray}
This may be computed by first integrating 
over $k_{||}$ employing
\begin{eqnarray}
\int_{-\infty}^{\infty} \frac{1-\cos[a(b-x)]}{x^{2}+c^{2}} dx = \frac{\pi}{c}[1-e^{-ac} \cos(ab)]
\end{eqnarray}
to obtain
\begin{eqnarray} 
\frac{1}{A_{||}}
\int_{0}^{\infty} \frac{1-\cos(k_{||}x_{||} +
 \vec{k}_{\perp} \cdot \vec{x}_{\perp})}
{k_{||}^{2}+ (A_{\perp} k_{\perp}^{4}/A_{||})} dk_{||} \nonumber
\\ = \frac{1}{2} \frac{\pi}{k_{\perp}^{2}}
\sqrt{\frac{1}{A_{||} A_{\perp}}}
[1- \exp(-\sqrt{\frac{A_{\perp}}{A_{||}}} \vec{k}_{\perp}^{2} x_{||})
\cos (\vec{k}_{\perp} \cdot \vec{x}_{||})]
\end{eqnarray}
If $\phi$ denotes the angle between $\vec{k}_{\perp}$ and 
$\vec{x}_{\perp}$ then
\begin{eqnarray}
\int_{0}^{2 \pi} 
[1-\exp(-\sqrt{\frac{A_{\perp}}{A_{||}}} \vec{k}_{\perp}^{2}x_{||}) 
\cos(\vec{k}_{\perp} \cdot \vec{x}_{\perp})] d \phi \nonumber
\\ = 2 \pi - \exp(-\sqrt{\frac{A_{\perp}}{A_{||}}} k_{\perp}^{2} x_{||})
\int_{0}^{2 \pi} \cos(k_{\perp} x_{\perp} \cos \phi) d \phi.
\end{eqnarray} 
As 
\begin{eqnarray}
J_{0}(x) = \frac{1}{\pi} \int_{0}^{\pi} \cos(x \cos \phi) 
d \phi,
\end{eqnarray}
\begin{eqnarray}
I(\vec{x}) = \frac{1}{2A_{||}} \pi 2 \pi \int_{0}^{\Lambda} 
\frac{1-\exp(-\sqrt{\frac{A_{\perp}}{A_{||}}} k_{\perp}^{2} x_{||})
J_{0}(k_{\perp} x_{\perp})}{\sqrt{\frac{A_{\perp}}{A_{||}}}
k_{\perp}^{2}} k_{\perp} dk_{\perp}.
\nonumber
\end{eqnarray}
We may now insert the series expansion of $J_{0}(x)$ and integrate
term by term.
\begin{eqnarray}
J_{0}(z) = \sum_{n=0}^{\infty} \frac{(-)^{n}z^{2n}}{2^{2n} (n!)^{2}}.
\end{eqnarray}
Comparing the result to the series form for the exponential
\begin{eqnarray}
E_{1}(z) = -\gamma - \ln z - \sum_{n=1}^{\infty} (-)^{n} \frac{z^{n}}{n(n!)}
\end{eqnarray}
($\gamma$ is Euler's constant) we find
that 
\begin{eqnarray}
G(\vec{x}) \sim \frac{4 d^{2}}{x_{\perp}^{2}}~ \exp[- 2 \eta \gamma -
\eta E_{1}(\frac{x_{\perp}^{2} q}{4 x_{||}
\sqrt{\frac{A_{\perp}}{A_{||}}}})]~\times ~ \cos[qx_{||}]
\end{eqnarray}
where $\eta= \frac{k_{B}T}{8 \pi} \sqrt{\frac{A_{||}}{A_{\perp}}}$
which in our case is $\frac{k_{B}T}{16 \pi}Q^{1/4}$ and 
$d= \frac{2 \pi}{\Lambda}$ 
where $\Lambda$ is the aforementioned  
ultra violet momentum cutoff.

\section{acknowledgments}

The bulk of this work
was covered in my thesis \cite{thesis}
two years ago. I wish to acknowledge
my mentors Steven Kivelson 
and Joseph Rudnick for allowing
students to pursue their 
own ideas. I am indebted
to many valuable conversations
with Lincoln Chayes, Steven A. Kivelson,
Joseph Rudnick, and James P. Sethna.


\begin{thebibliography}{10}
\bibitem{smart}
Note that even if we had not known the ground states
we could still prove that there is
no magnetization. 

All one would need to do is to 
apply an infinitesimal magnetic
field 
\begin{equation}
\vec{h}(\vec{x}) = h \vec{S}^{ground-state}(\vec{x})
\end{equation}
with as yet an unknown ground state $\vec{S}^{ground-state}(\vec{x})$ .

Replacing any appearance of the minimizing 
wave-vector $\vec{q}$ in the Schwarz inequality 
and the definition of $B(\vec{k})$ by the more general 
wave-vector $\vec{\ell}$ and setting
$h \to 0^{+}$ one would arrive at 
the conclusion that $m_{\vec{\ell}}=0$
for each mode $\vec{\ell}$. 
Consequently 
\begin{eqnarray}
\langle S_{n}(\vec{x}) \rangle = \frac{1}{N} \sum_{\vec{\ell}} e^{i
\vec{l} \cdot \vec{x}} m_{\ell} = 0
\end{eqnarray}

\bigskip


Stated alternatively, by Parseval's theorem,
\begin{eqnarray}
\sum_{\vec{x}} \langle \vec{S}(\vec{x}) \rangle^{2} = \frac{1}{N^{2}} 
\sum_{\vec{l}} \langle \vec{S}(\vec{l}) \rangle \langle 
\vec{S}(-\vec{l}) \rangle. 
\end{eqnarray}

Taking careful note of the system size ($N$) 
dependence in Eqn.(\ref{divergence1}) and Eqn.(\ref{main}), 
\begin{eqnarray}
\sum_{\vec{x}} \langle \vec{S}(\vec{x}) \rangle^{2} = {\cal{O}}
(\frac{N}{\ln N})
\end{eqnarray}
and thus the average value of $|\langle \vec{S}(\vec{x}) \rangle|$
at any given site $\vec{x}$ diminishes as
${\cal{O}}(\sqrt{1/\ln N})$
which vanishes in the thermodynamic ($ N \rightarrow \infty$)
limit.
\bibitem{explain}

Looking at the real space form of the constraining
term it is readily seen that $\langle \psi| H_{1} | \psi \rangle \ge 0$. 
If $|\delta S| \ll 1$ this implies that the quadratic term in $\delta
S(\vec{x})$ originating from $H_{1}$ is non-negative definite.
For $S_{i=3}^{g}(\vec{x}) =0$, this quadratic term in $\delta
S_{3}(\vec{x})$ vanishes. The lowest eigenvalue $\lambda_{\min} =2$
corresponds to this vanishing ${\cal{O}} (\delta S ^{2}) $
contribution from $H_{1}$. An analogous occurs 
in the fluctuation analysis about a poly-spiral state.
\end{thebibliography}
\end{document}